\documentclass[english,12pt,oneside]{article}
\pdfoutput=1
\usepackage[T1]{fontenc}
\usepackage[latin9]{inputenc}
\usepackage[usenames,dvipsnames]{xcolor}
\usepackage{color}
\usepackage{babel}
\usepackage{setspace}
\usepackage{amstext}
\usepackage{amssymb}
\PassOptionsToPackage{normalem}{ulem}
\usepackage{ulem}
\usepackage[unicode=true,pdfusetitle,
 bookmarks=true,bookmarksnumbered=false,bookmarksopen=false,citecolor=Turquoise,
 breaklinks=false,pdfborder={0 0 1},backref=false,colorlinks=true,pdfpagemode=FullScreen]
 {hyperref}
\usepackage{graphicx}
\usepackage{mathtools}
\usepackage[small]{caption}
\providecommand{\tabularnewline}{\\}

\definecolor{note_fontcolor}{rgb}{0.80078125, 0.80078125, 0.80078125}

\renewcommand{\url}[1]{\href{#1}{\textsc{\textcolor{Emerald}{WebSite}}}}

\long\def\symbolfootnote[#1]#2{\begingroup%
\def\thefootnote{\fnsymbol{footnote}}\footnote[#1]{#2}\endgroup}

\usepackage[top=3 cm, bottom=3 cm, left=3.5 cm, right=3.5 cm]{geometry}

\newenvironment{changemargin}[2]{%
\begin{list}{}{%
\setlength{\topsep}{0pt}%
\setlength{\leftmargin}{#1}%
\setlength{\rightmargin}{#2}%
\setlength{\listparindent}{\parindent}%
\setlength{\itemindent}{\parindent}%
\setlength{\parsep}{\parskip}%
}%
\item[]}{\end{list}}
\usepackage[]{caption}
\begin{document}

\begin{titlepage}
\begin{changemargin}{-2cm}{-2cm}
\vspace{.3in}
\hfill 
\vspace{2.5cm}
\begin{center}
\noindent \textsc{ \LARGE Resonance at 125 GeV: Higgs or Dilaton/Radion?
}
\end{center}

\begin{center}
\vspace{1.0cm}
{\bf Zackaria Chacko~\symbolfootnote[1]{\href{mailto:zchacko@umd.edu}
{zchacko@umd.edu} }\,\,, 
Roberto Franceschini~\symbolfootnote[2]{\href{mailto:rfrances@umd.edu}
{rfrances@umd.edu} }\,\,, 
and Rashmish K. Mishra~\symbolfootnote[3]{\href{rashmish@umd.edu}
{rashmish@umd.edu} }\,\,\\}
\setstretch{1.5}
Maryland Center for Fundamental Physics, Department of Physics,\\
University of Maryland, College Park, MD 20742, USA
\setstretch{1.}
\end{center}

\end{changemargin}

\vspace{.8cm}

\begin{abstract}
\medskip
\noindent

We consider the possibility that the new particle that has been observed 
at 125 GeV is not the Standard Model (SM) Higgs, but instead the dilaton 
associated with an approximate conformal symmetry that has been 
spontaneously broken. We focus on dilatons that arise from theories of 
technicolor, or from theories of the Higgs as a pseudo-Nambu-Goldstone 
boson (pNGB), that involve strong conformal dynamics in the ultraviolet. 
In the pNGB case, we are considering a framework where the Higgs 
particle is significantly heavier than the dilaton and has therefore not 
yet been observed. In each of the technicolor and pNGB scenarios, we 
study both the case when the SM fermions and gauge bosons are 
elementary, and the case when they are composites of the strongly 
interacting sector. Our analysis incorporates conformal symmetry 
violating effects, which are necessarily present since the dilaton is 
not massless, and is directly applicable to a broad class of models that 
stabilize the weak scale and involve strong conformal dynamics. Since 
the AdS/CFT correspondence relates the radion in Randall-Sundrum (RS) 
models to the dilaton, our results also apply to RS models with the SM 
fields localized on the infrared brane, or in the bulk. We identify the 
parameters that can be used to distinguish the dilatons associated with 
the several different classes of theories being considered from each 
other, and from the SM Higgs. We perform a fit to all the available data 
from several experiments and highlight the key observations to extract 
these parameters. We find that at present, both the technicolor and pNGB 
dilaton scenarios provide a good fit to the data, comparable to the SM 
Higgs. We indicate the future observations that will help to corroborate 
or falsify each scenario.
\end{abstract}

\end{titlepage}

\clearpage

\setstretch{1.5}

\setcounter{footnote}{0}
\setcounter{figure}{0}
\setcounter{table}{0}

\section{Introduction}

The Large Hadron Collider (LHC) is currently well on its way to 
uncovering the mechanism that underlies the breaking of electroweak 
symmetry. The discovery of a Higgs-like particle localized at a mass of 
about 125 GeV by ATLAS and CMS~\cite{The-CMS-Collaboration:2012ys,The-ATLAS-Collaboration:2012rt} is just the first 
step towards a thorough understanding of the phenomenon of electroweak 
symmetry breaking (EWSB).

The resonance observed at the LHC sits at a rather special spot for 
experiment. In particular, the decays of a SM Higgs boson with such a 
mass have sizable branching fractions into several final states 
including $b\bar{b},\, W^{+}W^{-},\,\tau\bar{\tau},\, ZZ$ and 
$\gamma\gamma$. Searching for signals of the new particle, produced 
either singly or in association with $W$ bosons, $Z$ bosons or well 
separated forward energetic jets, the LHC experiments expect to be able 
to measure with significant accuracy its branching ratios into each of 
these final states. These measurements will either confirm that the new 
state is the SM Higgs, or expose it as an impostor.

The availability of this data, and the presence of excesses above the 
expected background in several different channels (even above the SM 
Higgs signal hypothesis in some cases), has already motivated several 
analyses that undertake a global interpretation of the properties of the 
particle responsible for the new physics observed by ATLAS and CMS 
\cite{Carmi:2012yq,Azatov:2012rt,Espinosa:2012vn,Giardino:2012lr,Ellis:2012dq,Klute:2012yq,Corbett:2012rt,Giardino:2012qy,Carmi:2012vn,Espinosa:2012yq,Ellis:2012fj,Low:2012ys,Montull:2012kx,Buckley:2012fk,Gunion:2012uq}. 
Although constrained by the limited luminosity, these global analyses 
have yielded several interesting results. In particular, it has been 
shown that the 2011 data disfavors a fermio-phobic Higgs boson 
\cite{Gabrielli:2012rt,Gabrielli:2012vn}, and limits have been placed on 
the invisible decay width of the Higgs 
\cite{Giardino:2012lr,Espinosa:2012qy,Dreiner:2012lr}. Furthermore these 
efforts have led to a productive discussion among theorists and 
experimentalists about the way to present the results of experimental 
searches. As detailed in Refs.~\cite{Azatov:2012ys,Kraml:2012fr} some searches could  potentially have a wider reach in terms of physics scenarios that are probed. The wish 
 would be to see the analyses of inclusive and exclusive final states presented in a broad manner, as opposed to the (overly) tuned analysis performed so far in most cases
 
\bigskip

On the theoretical side, 125 GeV is also a rather special value for the 
SM Higgs boson mass. Such a value of the Higgs mass is in the far 
periphery of the preferred parameter space of the Minimal Supersymmetric 
Standard Model (MSSM), and motivates the study of alternatives to 
supersymmetry that stabilize the weak scale. Several well-motivated 
models of electroweak symmetry breaking that solve the hierarchy problem 
involve strong conformal dynamics above the weak scale. In particular, 
in models of technicolor~\cite{Susskind:1978ms, Weinberg:1975gm}, for a 
review see~\cite{Hill:2002ap}, and of the Higgs as a 
pseudo-Nambu-Goldstone boson (pNGB)~\cite{Georgi:1974yw,Georgi:1975tz, 
Kaplan:1983fs, Kaplan:1983sm,Georgi:1984af}, strong conformal dynamics 
can help separate the flavor scale from the weak 
scale~\cite{Luty:2006sp}, (see also~\cite{Holdom:1984sk, 
Appelquist:1986an, Yamawaki:1985zg, 
Appelquist:1986tr,Appelquist:1987fc}), allowing the bounds on 
flavor-changing neutral currents to be satisfied.

In theories where conformal symmetry is spontaneously broken, the low 
energy effective theory contains a massless scalar, the dilaton, which 
may be thought of as the Nambu-Goldstone boson (NGB) associated with the 
breaking of conformal symmetry~\cite{Salam:1970qk, Isham:1971dv, 
Zuminolectures, Ellis:1970yd}. The form of the dilaton couplings is 
fixed by the requirement that conformal symmetry be realized 
non-linearly, and so this framework is extremely predictive. Several 
authors have studied the self-interactions of a light dilaton, and its 
couplings to SM fields ~\cite{Rattazzi:2000hs, Goldberger:2007zk, 
Vecchi:2010gj,Chacko:2012sy}. Interestingly, the couplings of a 
dilaton to the SM gauge and matter fields are very similar to those of a 
SM Higgs~\cite{Goldberger:2007zk}. The reason is that at the classical 
level the SM has an approximate conformal symmetry which is violated 
only by the Higgs mass term. Since the VEV of the Higgs doublet 
spontaneously breaks this approximate conformal symmetry, in the 
classical limit the couplings of the Higgs to the SM fields are very 
similar to those of a dilaton. It is therefore very natural to ask the 
question whether the resonance observed by the LHC at 125 GeV could be 
the dilaton associated with a strongly interacting conformal sector that 
breaks electroweak symmetry, rather than the SM Higgs. This is the 
question we shall focus on in this paper.

The remarkable fact that a dilaton can naturally mimic the SM Higgs 
boson has already lead to significant interest in the problem of 
distinguishing the two at colliders~\cite{Goldberger:2007zk,Fan:2008jk,Coleppa:2012wd,Campbell:2012lr,de-Sandes:2012gf,Barger:2012qf,Barger:2012ve}. 
As such it is not surprising that the idea that the new physics behind 
the 125 GeV resonance could be the dilaton (in combination or not with a 
Higgs boson) has already been put 
forward~\cite{Elander:2012fv,Matsuzaki:2012rp,Matsuzaki:2012fr,Matsuzaki:2012pd,Matsuzaki:2012pi,Grzadkowski:2012bh,Cheung:2012lq}, \cite{Giardino:2012lr,Ellis:2012dq,Giardino:2012qy,Carmi:2012vn,Espinosa:2012yq,Ellis:2012fj,Low:2012ys}. 
However the types of dilaton that have been considered so far in the 
post-LHC literature are rather specific, and the analyses do not apply 
to many of the theories that are of of primary interest for electroweak 
symmetry breaking. Furthermore, these studies do not include the effects 
of conformal symmetry violation, which is necessarily present since the 
dilaton is not massless, and can sometimes be significant. Our analysis 
extends to more general incarnations of the dilaton, incorporates 
conformal symmetry violating effects, and is directly applicable to a 
broad class of models that stabilize the weak scale and involve strong 
conformal dynamics.

Specifically we consider the possibility that the new particle that has 
been observed at 125 GeV is a dilaton that arises from a theory of 
technicolor, or from a theory of the Higgs as a pNGB, that involves 
strong conformal dynamics in the ultraviolet. In the pNGB case, we are 
considering a framework where the Higgs particle is significantly 
heavier than the dilaton and has therefore not yet been observed. In 
each of the technicolor and pNGB scenarios, we study both the case when 
the SM fermions and gauge bosons are elementary, and the case when they 
are composites of the strongly interacting sector. We identify the 
parameters that can be used to distinguish the dilatons associated with 
the several different classes of theories being considered from each 
other, and from the SM Higgs. We perform a fit to all the available data 
from several experiments and highlight the key observations to extract 
these parameters. We also indicate the future observations that should 
help to corroborate or falsify each scenario.

The AdS/CFT correspondence~\cite{Maldacena:1997re, Witten:1998qj, 
Verlinde:1999fy, Verlinde:1999xm} relates Randall-Sundrum (RS) models in 
warped extra dimensions~\cite{Randall:1999ee} to strongly coupled 
conformal field theories in the large $N$ limit. In this way, extra 
dimensional realizations of technicolor~\cite{Csaki:2003zu} and of the 
Higgs as a pNGB~\cite{Contino:2003ve,Agashe:2004rs} have been obtained. 
In the correspondence, the radion in the RS model is identified with the 
dilaton~\cite{Rattazzi:2000hs}. The couplings of the radion to the SM 
gauge and matter fields have been studied, both in the case when the SM 
fields are localized to the infrared brane~\cite{Goldberger:1999uk,Csaki:2000lr, 
Goldberger:1999un,Giudice:2000av, Csaki:2000zn} and in the case when 
they are in the bulk~\cite{Rizzo:2002pq, Csaki:2007ns}, and agree with 
the dilaton couplings in the appropriate limit. Our results on the 
dilaton therefore also apply to the radion in RS models, with the SM 
fields localized on the brane, or in the bulk.

\section{Interactions of the dilaton}

\subsection{General Considerations}

The fifteen parameter conformal group includes scale transformations and 
special conformal transformations, in addition to the ten parameters of 
the Poincare group. If conformal invariance is spontaneously broken the 
low energy effective theory contains a massless dilaton field 
$\sigma(x)$, which can be thought of as the NGB associated with the 
breaking of scale invariance~\cite{Salam:1970qk, 
Isham:1971dv,Zuminolectures, Ellis:1970yd,Isham:1970gz,Ellis:1971sa}.  
The additional four NGBs associated with the breaking of the special 
conformal symmetry can be identified with the derivatives of the 
dilaton, rather than as independent dynamical fields. Below the breaking 
scale the symmetry is realized non-linearly, with the dilaton undergoing 
a shift $\sigma(x) \rightarrow \sigma'(x') = \sigma(x) + \omega f$ under 
the scale transformation $x^{\mu} \rightarrow x'^{\mu} = e^{- \omega} 
x^{\mu}$. Here $f$ is the scale associated with the breaking of 
conformal symmetry. For the purpose of writing interactions of the 
dilaton it is convenient to define the conformal compensator
 \begin{equation}
\chi(x) = f e^{\sigma(x)/f}
 \end{equation}
which transforms linearly under scale transformations. Specifically,
under the scale transformation $x^{\mu} \rightarrow x'^{\mu} =
e^{-\omega} x^{\mu} $, $\chi(x)$ transforms as
 \begin{equation}
\chi(x) \rightarrow \chi'(x') = e^{\omega} \; \chi(x) \; .
\label{dilatontransformation}
 \end{equation}
The low energy effective theory for the dilaton will in general include 
all terms consistent with this transformation, but with some additional 
restrictions on their coefficients from the requirement that the theory 
be invariant not just under scale transformations, but under the full 
conformal group. These restrictions will not affect our discussion in 
any significant way, and so in practice we shall only require that the 
action for $\chi$ be scale invariant.

In realistic models of electroweak symmetry breaking conformal symmetry 
is not exact. As a consequence the dilaton will acquire a mass, and 
there will be corrections to its couplings from conformal symmetry 
violating effects. If the dimension of the operator responsible for the 
breaking of conformal symmetry is close to marginal even at the breaking 
scale, the mass of the dilaton can naturally lie below the scale of 
strong dynamics~\cite{RattazziPlanck2010, Chacko:2012sy}, see 
also~\cite{Appelquist:2010gy}. Although this condition is, in general, 
not satisfied in the scenarios of interest for electroweak symmetry 
breaking, the presence of a light dilaton in these theories is only 
associated with mild tuning~\cite{Chacko:2012sy}. This makes it a 
priori quite plausible that the Higgs-like particle observed at the LHC 
is in fact a dilaton. Corrections to the form of the dilaton couplings 
to SM states from the conformal symmetry violating effects that generate the dilaton mass are suppressed by the square of the ratio of the dilaton 
mass to the strong coupling scale~\cite{Chacko:2012sy}. Although this 
is a small parameter in the theories of interest, corrections to the 
couplings can nevertheless sometimes be significant and will need to be 
taken into account in our analysis.

The strength of the dilaton couplings is controlled by the parameter 
$f$, the scale at which conformal symmetry is broken. In technicolor 
models, the same condensate that breaks electroweak symmetry also breaks 
conformal symmetry. Hence we expect that $f$ is of order the electroweak 
VEV, $v$, in theories of technicolor. For models of the Higgs as a pNGB, 
however, the situation is different. It is the same dynamics that breaks 
the global symmetry of which the Higgs is a pNGB that now breaks 
conformal symmetry. Then, depending on the scale at which the global 
symmetry is broken, there can be a hierarchy between $v$ and $f$. Since 
precision electroweak measurements prefer the Higgs compositeness scale 
$\Lambda \sim 4 \pi f$ to be 5 TeV or larger, the favored values of $f$ 
are such that $f \gtrsim 3v$.

We introduce the variable
 \[
\xi=\frac{v^{2}}{f^{2}}
 \]
to parametrize the ratio of breaking scales. Technicolor theories are 
associated with values of $\xi$ close to one. Meanwhile $\xi=0$ 
describes a pNGB-like model with the cutoff pushed very high, 
corresponding to the SM-like limit of these theories. In this limit, the 
pNGB scenario is highly tuned and completely fails as a solution to the 
hierarchy problem. On the other hand precision electroweak measurements 
prefer the Higgs compositeness scale $\Lambda \sim 4 \pi f$ to be 5 TeV 
or larger. As such a moderate $\xi$ is expected in pNGB models to solve 
the hierarchy problem and not be in tension with precision tests.

\subsection{Dilaton couplings to the SM Fields}

The next step is to understand the dilaton couplings to the SM fields. 
Mixing between the dilaton and a pNGB Higgs is generally small, 
particularly if $v \ll f$ or if the dilaton is significantly lighter 
than the Higgs (or vice versa). We shall therefore neglect it. The form 
of the dilaton couplings to the SM fermions and gauge bosons is then the 
same in the technicolor and pNGB cases~\cite{Chacko:2012sy}.

We will consider in turn the dilaton couplings to massive gauge bosons, 
massless gauge bosons and fermions. The form of these interactions 
depends on whether the SM fields are elementary particles or composites 
of the strongly interacting sector, and we will consider both 
possibilities. The AdS/CFT correspondence relates composite fermions and 
gauge bosons to states localized on or towards the infrared brane in RS 
models. Elementary fermions are associated with states localized towards 
the ultraviolet brane and elementary gauge bosons live in the bulk of 
the space. The radion couplings to brane and bulk fields in RS models 
can therefore be determined from the corresponding interactions of the 
dilaton. Our results for the dilaton are therefore directly applicable 
to the radion in certain classes of RS models. However, in our analysis we will not consider constraints on the parameters of the RS model rising from 
limits on other particles in the spectrum, such as Kaluza-Klein excitations of the graviton, (see for example \cite{Tang:2012ly,Tang:2012yq}).

\subsubsection{Dilaton couplings to the massive gauge bosons}

Depending on the model, the massive gauge bosons of the SM, the $W$ and 
$Z$, could either be elementary particles or composite states that 
emerge at low energies from the strong conformal dynamics. In either 
case the gauge boson mass term 
 \[ \frac{m_{W}^{2}}{g^{2}}W^{2} 
 \] 
is the leading effect which breaks conformal symmetry. Then the dilaton 
couples in such a way as to compensate for this~\cite{Goldberger:2007zk, 
Chacko:2012sy},
 \[ 
\frac{\chi^2}{f^2} \frac{m_{W}^{2}}{g^{2}}W^{2} 
\rightarrow 2 \frac{\sigma}{f} \frac{m_{W}^{2}}{g^{2}}W^{2} \,.
 \]
Note that we are working in a basis where the gauge boson kinetic terms
have the form 
 \[
-\frac{1}{4g^{2}}F_{\mu\nu}F^{\mu\nu} \,.
 \]
We see that the coupling of the dilaton is in general suppressed by a 
factor $\frac{v}{f} = \sqrt{\xi}$ with respect to the coupling of a SM 
Higgs boson. Corrections to the form of this interaction from conformal
symmetry violating effects are small~\cite{Chacko:2012sy}.

\subsubsection{Dilaton coupling to the massless gauge bosons}

In the case of the massless gauge bosons, the photon and the gluon,
there is no classical source of breaking of the conformal symmetry.
However, at the quantum level conformal invariance is broken, as
manifested by the renormalization group evolution of the gauge couplings
\[
\frac{d}{d\log\mu } \frac{1}{g^{2}} = \frac{b}{8\pi^{2}} \,.
\]
The dilaton couples in such a way as to compensate for this effect.
The form of this coupling depends on whether the gauge bosons are
elementary particles or composites of the conformal field theory.

Let us consider first the case when the photon and gluon are elementary. 
Then the evolution of the gauge couplings is different above and below 
the strong coupling scale $\Lambda \sim 4 \pi f$. We parametrize the 
running of the gauge couplings above and below $\Lambda$ by $b_{>}$ and 
$b_<$ respectively. Above the top mass, $b_< = +7$ for the gluon and 
$-11/3$ for the photon. The running of the gauge couplings above the 
scale $\Lambda$ constitutes an explicit, rather than spontaneous, 
breaking of conformal symmetry. A spurion analysis then shows that the 
dilaton couples to elementary massless gauge bosons 
as~\cite{Chacko:2012sy}
 \begin{equation}
\label{dil2elementaryphoton}
\frac{1}{32\pi^{2}}\left(b_{<}-b_{>}\right)\frac{\sigma}{f}
F_{\mu\nu}F^{\mu\nu}\,.
 \end{equation}
This formula is valid at scales slightly below the strong coupling scale 
$4 \pi f$. Corrections from the conformal symmetry violating effects that generate the dilaton mass are 
again small~\cite{Chacko:2012sy}.

In the class of theories of interest the conformal sector necessarily 
transforms under electromagnetism. However in theories where the top 
quark is not composite, this sector need not be charged under the SM 
color group. Then $b_> = b_<$ for SM color, and the leading coupling of 
the dilaton to the gluons arises from a calculable top loop. In theories 
where the top quark is composite the conformal sector is necessarily 
charged under the SM color group.  However, since the gluon is 
elementary, $b_{>}$ is not constrained by conformal invariance, and its 
value depends on details of the specific conformal sector at hand. The 
difference $b_{>} - b_{<}$ is associated with the number of states in 
the strongly interacting sector with masses of order $\Lambda$, and is 
expected to be of order a few.

\bigskip

In the case of composite gauge bosons, there is no explicit breaking of 
conformal symmetry above the scale $\Lambda$.  As a consequence, the 
formula Eq.~(\ref{dil2elementaryphoton}) still applies, but with $b_> = 
0$~\cite{Goldberger:2007zk}. Since $b_<$ is known, it would appear that 
this framework is very predictive. However, in this scenario, 
corrections to the dilaton couplings from conformal symmetry violating 
effects can be significant. These take the form~\cite{Chacko:2012sy}
 \[
\frac{c}{4 g^2} \frac{m_{\sigma}^2}{\Lambda^2}
\frac{\sigma}{f}
F_{\mu\nu}F^{\mu\nu}\,,
  \]
where $c$ is of order one. For $\Lambda$ of order the TeV scale, and 
$m_{\sigma}$ of order 100 GeV, this gives a contribution to the dilaton 
couplings comparable to Eq.~(\ref{dil2elementaryphoton}), and 
predictivity is lost. This is particularly so in the case of the photon, 
less so for the gluon. For larger $\Lambda \gtrsim 3$ TeV, these 
corrections are much smaller, and predictivity is restored. It follows 
from this that the couplings of the dilaton to massless gauge bosons, if 
composite, differ from those of the SM Higgs by the product of 
$\sqrt{\xi}$ and a multiplicative factor of order a few. This factor can 
be predicted for larger values of $\Lambda$, but not for small 
$\Lambda$.

\subsubsection{Dilaton couplings to the SM fermions \label{fermioncouplings}}

In the case of the SM fermions the leading source of conformal symmetry 
breaking are the fermion mass terms. The coupling of the dilaton depends 
on whether the SM fermions are purely elementary particles or are 
composites, or partial composites of the strong conformal dynamics.

\bigskip

In the case of elementary fermions their mass arises from direct 
couplings to a scalar operator $\mathcal{H}$ in the conformal field 
theory with the gauge quantum numbers of the SM Higgs, as in theories of 
conformal technicolor~\cite{Luty:2006sp}. For the third generation 
up-type quarks these couplings take the form
 \[
\mathcal{L}=y_{top}Q_{3}\mathcal{H}t^c + h.c.\,\,.
 \]
This leads to the mass term for the top which we write as 
 \[
m_t Q_{3} t^c \,.
 \]
The condition that the flavor problem be addressed means that the large 
value of the top mass constrains the size of $\Delta_{\mathcal{H}}$. If 
we denote the flavor scale by $\Lambda_{\rm F}$, we require
 \[
\left(\frac{\Lambda_{\rm F}}{4 \pi f} \right)^{\Delta_{\mathcal{H}}-1}
\lesssim \frac{4\pi v}{m_t} \,.
 \]

For $\Lambda_{\rm F} \gtrsim 1000$ TeV, we require 
$\Delta_{\mathcal{H}}\lesssim 1.3 $ for the technicolor case. For the 
pNGB case, with $f \sim$ 1 TeV, the constraint becomes 
$\Delta_{\mathcal{H}} \lesssim 1.5 $.  At the same time, the hierarchy 
problem constrains the dimension of the operator $\mathcal{H}^{\dagger} 
\mathcal{H}$ to be of order 4 or larger. Since $\Delta_{\mathcal{H}} = 
1$ necessarily implies that the conformal field theory is free, which in 
turn implies $\Delta_{\mathcal{H}^{\dagger} \mathcal{H}} = 2$ , the 
conditions $\Delta_{\mathcal{H}} \lesssim 1.3$ (or $\Delta_{\mathcal{H}} 
\lesssim 1.5$) and $\Delta_{\mathcal{H}^{\dagger}\mathcal{H}} \gtrsim 4$ 
are in tension. Note that these conditions cannot be simultaneously 
satisfied in the large $N$ limit, and therefore realistic models of this 
type with elementary fermions cannot be constructed within the RS 
framework. Unitarity and causality can be used to constrain 
$\Delta_{\mathcal{H}} \gtrsim 1.5$ if 
$\Delta_{\mathcal{H}^{\dagger}\mathcal{H}} \gtrsim 
4$~\cite{Rattazzi:2008pe,Rychkov:2009ij,Rattazzi:2010yc},~\cite{Vichi:2011ux}, 
~\cite{Poland:2011ey}, (see also~\cite{Poland:2010wg}). Therefore the 
technicolor scenario with elementary fermions is somewhat tuned.

A spurion analysis shows that the dilaton couples to the top quarks 
as~\cite{Vecchi:2010gj, Chacko:2012sy}
 \[
\frac{m_t}{f}\Delta_{\mathcal{H}}\sigma Q_{3}t^c\,.
 \]
This differs by a factor of $\sqrt{\xi}\Delta_{\mathcal{H}}$ from the top
Yukawa coupling in the SM. More generally, the couplings of all the fermions
to the dilaton will differ from that of the SM Higgs by this factor.
Corrections to this expression from the conformal symmetry violating effects that lead to the dilaton mass
are small~\cite{Chacko:2012sy}. 

From this discussion it is clear that a useful parametrization of the
coupling of the dilaton to the top quark in the case of elementary
fermions is as
 \[
\frac{m_{t}}{f}\left(1+\epsilon\right)\sigma Q_3 t^c\,,
 \]
where $\epsilon=\Delta_{\mathcal{H}}-1$.

More generally, we can apply the same parametrization to the dilaton 
couplings with the other fermions as well, so that the interactions of 
the dilaton with elementary SM fermions differ from those of the Higgs 
by a universal factor of $\sqrt{\xi} (1 + \epsilon)$.

\bigskip

Another possibility is that the SM fermions are partially composite, 
emerging from the mixing of elementary fermions with operators in the 
conformal field theory that have the same gauge quantum 
numbers~\cite{Kaplan:1991dc, Contino:2006nn}. For the up-type quarks 
this takes the form
 \begin{equation}
\label{compositeup}
\mathcal{L}=\tilde{y}_{Q}Q\mathcal{Q}+\tilde{y}_{u}u^c \mathcal{U}+h.c.\,\,.
 \end{equation}
Here $\mathcal{Q}$ and $\mathcal{U}$ are operators in the CFT with 
scaling dimension $\Delta_{Q}$ and $\Delta_{u}$, respectively, and for 
simplicity we are suppressing flavor indices. The resulting mass term is 
of the form 
\[ m_{u}Q u^c\,,
\]
where $m_{u}$ is proportional to the 
product $\tilde{y}_{Q}\tilde{y}_{u}$ up to effects that are higher order in 
$y$. A spurion analysis then show that the dilaton couples 
as~\cite{Vecchi:2010gj, Chacko:2012sy}
 \[
\frac{m_{u}}{f}\left(\Delta_{Q}+\Delta_{u}-4\right)\,\sigma \, Qu^c\,.
 \]
We see that this coupling differs from that of the SM Higgs boson by the 
factor $\frac{v}{f}\left(\Delta_{Q} + \Delta_{u}-4\right)$. Corrections 
from the conformal symmetry violating effects that generate the dilaton mass are 
small~\cite{Chacko:2012sy}.

However, as in the case of elementary fermions, the scaling dimensions 
of the relevant operators are bound by phenomenological considerations. 
In fact to generate a large mass for the top quark we need 
$\tilde{y}_{Q}$ and $\tilde{y}_{u}$ for the third generation to be of 
order one at the scale $\Lambda$. For a high flavor scale $\Lambda_{\rm 
F}$, in the absence of tuning, this implies that the terms in 
Eq.~(\ref{compositeup}) that generate the top quark mass are close to 
marginal, corresponding to $\Delta_{Q}\simeq 5/2$ and 
$\Delta_{u}\simeq 5/2 \,\,$. As such in this scenario the deviation from 
the coupling of the SM Higgs is determined by the extent to which the 
terms in Eq.~(\ref{compositeup}) deviate from exact marginality.

For $\Delta_{Q} = \Delta_{u} = 5/2$, the terms in 
Eq.~(\ref{compositeup}) are exactly marginal and do not violate 
conformal symmetry. In this limit the coupling of the dilaton is 
suppressed simply by a factor of $\sqrt{\xi} = v/f$ with respect to the 
coupling of the SM Higgs boson. Not surprisingly, the same result is 
obtained if the SM fermions $Q$ and {$u^c$} are fully composite states, 
since in this limit the Yukawa coupling again arises from an effect that 
does not violate conformal symmetry.

For our analysis, the dilaton couplings to the third generation fermions 
are particularly important, since the current data is only sensitive to 
these interactions. For concreteness, in our analysis of partially 
composite fermions, we will set $\Delta_{Q} = \Delta_{u} = 5/2$ for the 
top quark, corresponding to a fully composite top. More generally, we 
will assume that all the third generation SM fermions are primarily 
composite, so that their coupling to the dilaton is identical to that of 
the SM Higgs up to an overall factor of $\sqrt{\xi}$.

\section{Global fit of the Higgs data }

\subsection{Couplings and relevant observables}

\begin{table}
\begin{centering}
$\begin{array}{ccc}
 & \textrm{{\bf Elementary   }} & \textrm{{\bf Composite}}\\
 & \textrm{{\bf Fermions   }} & \textrm{{\bf  Fermions}}\\

\\
\eta_{\text{bb}} & (\epsilon+1)^{2}\xi & \xi\\
\eta_{\text{WW}} & \xi & \xi\\
\eta_{\tau\tau} & (\epsilon+1)^{2}\xi & \xi\\
\eta_{\text{ZZ}} & \xi & \xi\\
\eta_{\text{gg}} & (\epsilon+1)^{2}\xi & \xi\psi\\
\eta_{\gamma\gamma} & \xi\phi & \xi\phi
\end{array}$
\par\end{centering}

\caption{\label{tab:couplings}Summary of the couplings of the dilaton.}
\end{table}

In this section we summarize the couplings of the general dilaton 
introduced earlier, and identify the observables in the Higgs searches 
that are particularly sensitive to the parameters of the theory. We 
specifically focus on the three different classes of theories below:
 \begin{itemize}
 \item
Theories where the SM fermions and gauge bosons are all elementary,
 \item
Theories where the SM matter fields are composites of the strong
dynamics, but the gauge fields are elementary.
 \item
Theories where the SM matter and gauge fields are all composite.
 \end{itemize}
For a more compact expression of the parametric behavior of the event 
rate we introduce the ratio of the square of the coupling of the dilaton 
to a given SM state over the square of the corresponding Higgs coupling 
in the SM,
 \[
\eta_{XX}\equiv\left(\frac{g_{\sigma XX}}{g_{hXX}}\right)^{2}\,.
 \]
The values of $\eta$ at tree-level associated with the different SM 
states are shown in Table~\ref{tab:couplings}, both for the case when 
the third generation fermions are elementary and the case when they are 
composite. In the scenario where the third generation fermions are 
composite, the SM gauge bosons could be either elementary or composite. 
We parametrize both cases the same way because the only significant 
difference in the analysis arises from the fact that if the SM gauge 
bosons are composite, then for larger values of $f$, corresponding to 
the range $\xi \lesssim 1/10$, the values of the parameters $\phi$ and 
$\psi$ are fixed.

We also remark that our parametrization of the couplings to massless 
vectors does not distinguish between calculable contributions of SM 
states and a priori unknown and largely unconstrained contributions 
arising from CFT states. This parametrization has the advantage of 
simplicity and can be easily mapped onto the predictions of specific UV 
scenarios. For instance in the case of composite gauge bosons the 
prediction would be:
   \begin{equation}
 \psi \simeq 132\,, \quad \phi \simeq 2.4   \,. \label{psi0phi0}
   \end{equation}
It is also possible to parametrize the coupling to massless gauge bosons 
in a way that separates the SM and CFT contributions. Since the SM 
contribution can be calculated, an experimental determination of the 
dilaton couplings to the massless gauge bosons would in principle allow 
the possibility of measuring properties of the CFT (up to the remaining 
uncertainty on the SM part). Since our primary focus is on the issue of 
distinguishing the SM Higgs from the dilaton, we leave this for future 
work.
    
In what follows we consider in turn the scenarios with elementary 
fermions and composite fermions, and identify in each case the 
observations that will allow the parameters in Table~\ref{tab:couplings} 
to be extracted.

\subsubsection{Elementary Fermions}

For elementary SM fermions the 
rate for the final state $X\bar{X}$ in the case of a dilaton produced 
from gluons through a top loop is given by
 \begin{eqnarray}
\sigma_{GF}^{(X\bar{X})} & = & \sigma_{GF}
\frac{\Gamma(\sigma\to X\bar{X})}{\Gamma_{\sigma}} \nonumber \\
& \simeq & \eta_{X\bar{X}}\sigma_{GF,\, SM}^{(X\bar{X})}
\left\{ 1+O\left(1-\frac{\Gamma(\sigma\to bb)}{\Gamma_{\sigma}}\right)\right\}  \nonumber \\
& \simeq &  \eta_{X\bar{X}}\sigma_{GF,\, SM}^{(X\bar{X})},
\label{eq:GFrate}
 \end{eqnarray}
where by $\sigma_{GF,\, SM}^{(X\bar{X})}$ is meant the Gluon Fusion (GF)
cross-section for a SM Higgs decaying into $X\bar{X}$. The last line 
follows from the fact that the dilaton decay width, like that of the 
Higgs, is dominated by the $b \bar{b}$ final state. 

The final states most sensitive to a Higgs-like dilaton produced through 
gluon fusion are $\gamma\gamma$, $WW^{*}$ and $ZZ^{*}$. These processes 
are insensitive to $\epsilon$, but depend on $\xi$, and in the case of 
the $\gamma \gamma$ final state, also on $\phi$. The effect of 
$\epsilon$ in the total width is largely cancelled by the enhancement in the 
coupling to the top that generates the coupling of the dilaton to the 
gluons. The fact that the observed rate to $ZZ^{*}$ is comparable to 
that of the SM Higgs suggests that in this scenario values of $\xi$ 
very different from unity are disfavored.

\bigskip

A similar analysis for the Vector~Boson~Fusion~(VBF) and 
Associated~Production~(AP) modes leads to
 \begin{eqnarray}
\sigma_{VBF,\, AP}^{(X\bar{X})} & = & \sigma_{VBF,\, AP}
\frac{\Gamma(\sigma\to X\bar{X})}{\Gamma_{\sigma}} \nonumber \\
 & \simeq & \xi\frac{\eta_{X\bar{X}}}{\eta_{bb}}
\sigma_{VBF,AP,\, SM}^{(X\bar{X})}\,.
\label{eq:VBFAPrate}
 \end{eqnarray}
We see that in contrast to gluon fusion, the rates of VBF and AP 
processes do depend on $\epsilon$, except in the case of the $b \bar{b}$ 
and $\tau^+ \tau^-$ final states. 

To isolate the effect of $\epsilon$ it is useful to take the ratio of GF 
and VBF (or AP) rates to the same final state for the dilaton, and for the
SM Higgs. Then,
 \begin{equation} 
\frac{    \sigma_{GF}^{(X\bar{X})} /  \sigma_{GF,\,SM}^{(X\bar{X})}  }
{  \sigma_{VBF,AP}^{(X\bar{X})}  / \sigma_{VBF,AP,\,SM}^{(X\bar{X})  } }
=\left(1+\epsilon\right)^{2}\,.
\label{eq:epsilonSensitive} 
 \end{equation}
We remark that, especially for the measurement of a small $\epsilon$, it might be relevant to include higher-order corrections to the couplings under study, which, in general, make this ratio a function of all the modified couplings of this scenario.

We see from this analysis that the parameters $\xi$, $\epsilon$ and 
$\phi$ can all be independently determined by combining different 
channels.

\subsubsection{Composite Fermions}

For composite SM fermions the parameter $\psi$ replaces $\epsilon$ in the parametrization of the coupling of the dilaton to the gluons.
We first consider the case of 
dilatons produced through gluon fusion,
 \begin{equation}
\sigma_{GF}^{(X\bar{X})}
\simeq   \eta_{X\bar{X}} \psi \, \sigma_{GF,\, SM}^{(X\bar{X})} \,.
 \end{equation}
We see that the rates to $\gamma\gamma$, $WW^{*}$ and $ZZ^{*}$ now
depend on both $\psi$ and $\xi$, and in the case of $\gamma \gamma$
on $\phi$ as well. 

For dilatons produced through vector boson fusion or associated production,
we have
 \begin{equation}
\sigma_{VBF,\, AP}^{(X\bar{X})} \simeq  \eta_{X\bar{X}} \,  \sigma_{VBF,AP,\, SM}^{(X\bar{X})}\,.
 \end{equation}
It follows that the VBF and AP modes are sensitive to $\xi$, but not to
$\psi$. 

In analogy with Eq.~(\ref{eq:epsilonSensitive}), one can isolate 
the effect of $\psi$ by taking the ratio 
 \begin{equation}
\frac{  \sigma_{GF}^{\left(X\bar{X}\right)}  /    \sigma_{GF,SM}^{\left(X\bar{X}\right)} } {    \sigma_{VBF,AP}^{(X\bar{X})}  /  \sigma_{VBF,AP,SM}^{(X\bar{X})}   }
=\psi\,.
\label{eq:psiSensitive} 
 \end{equation}
Once again we see that the parameters $\xi$, $\phi$ and $\psi$ can all 
be independently determined by appropriately combining the results of 
different channels.

\subsection{Distinguishing Elementary and Composite Fermions}

The scenarios with composite and elementary fermions correspond to very 
different theories in the ultraviolet (UV). Despite these stark 
differences in the UV distinguishing them on the basis of low energy 
observations might not be simple. For instance the couplings of the 
dilaton to gluons in the two scenarios are related by a change of 
parameters
 \begin{equation}
\psi \leftrightarrow (1+\epsilon)^{2}\,.
\label{epsilonpsi}
 \end{equation}
As the coupling to vector boson is the same it follows that from the
study of the production cross-sections in the gluon fusion and vector
boson fusion modes we cannot distinguish the two scenarios~\footnote{The
only restriction that applies in the relation Eq.~(\ref{epsilonpsi}) is
that $\epsilon>0$, hence when $\psi<1$ the scenario with elementary
fermions cannot be mapped onto the composite fermions scenario.}. Some
more information might be obtained by studying the (rare) production in
association with $t$ or $b$ quarks or by adding to the analysis the
information from the decay rates.

The key quantity to distinguish the two scenarios is the ratio 
$$\eta_{gg}/\eta_{bb}\quad \textrm{ or } \quad 
\eta_{gg}/\eta_{\tau\tau}\,,$$ which is unity~\footnote{Strictly 
speaking $\eta_{gg}/\eta_{bb}$ is one in the elementary fermions 
scenario only at tree-level. Going to higher order it becomes a function 
of all the other parameters. Furthermore both QCD and EW corrections 
should be included. In general it remains true that for the elementary 
fermions this ratio is predictable and fixed in terms of the other 
parameters, which makes the scenario distinguishable from the case with 
composite fermions.} in the elementary top scenario and a free 
parameter in the composite top scenario. To extract this quantity we can 
take the ratio of two generic $GF$ and $VBF$ (or $AP$) modes $X\bar{X}$ and 
$Y\bar{Y}$ in units of the SM rates
 \begin{equation}
\frac{ \sigma_{GF}^{(X\bar{X})} / \sigma_{GF,\,SM}^{(X\bar{X})}}{
\sigma_{VBF,AP}^{(Y\bar{Y})} / \sigma_{VBF,AP,\,SM}^{(Y\bar{Y})} } =
\frac{\eta_{X\bar{X}} }{\eta_{Y\bar{Y}}} \frac{\eta_{gg}}{\eta_{WW}}
\label{genericmuratio}
 \end{equation}
and focus on the case $X\bar{X}=WW$, $Y\bar{Y}=b\bar{b}$ to obtain
 \begin{equation}
\frac{ \sigma_{GF}^{(WW)} / \sigma_{GF,\,SM}^{(WW)}}{
\sigma_{VBF,AP}^{(b\bar{b})} / \sigma_{VBF,AP,\,SM}^{(b\bar{b})} } =
\frac{\eta_{gg}}{\eta_{bb}} \,.
 \end{equation}
Analogously one can study the $\tau^+ \tau^-$ final state to get further
information.

At the present time the errors on the measurements are still too large 
to draw conclusions on the value of this ratio. However we remark that 
both ATLAS and CMS reported measurements of the cross-sections involved, 
steadily improving their results. As such if the dilaton interpretation 
of the 125 GeV boson will turn out to be more appropriate than the one 
in terms of a SM Higgs boson, there is a chance that the measured 
cross-sections will provide a handle on the presence of colored states 
in the CFT in the UV.

\subsection{Experimental data}

The LHC and TeVatron experiments have searched for Higgs-like particles 
in a variety of final states. The data analyzed so far exhibits 
significant excesses over the expectation for the SM background. 
Striking evidence for a new boson has conclusively emerged from several 
analyses ~\cite{ATLAS-CONF-2012-093,CMS-PAS-HIG-12-020}.

While these observations still need to be refined the picture that has 
emerged seems to point towards new physics associated with a narrow 
resonance of mass about 125 GeV. As such the experiments tried to test 
an interpretation of the new boson in terms of a SM Higgs boson. To do 
that, for each relevant searched production and decay mode they presented the best 
fit value of the \emph{signal strength} of a SM Higgs boson
\[
\bar{\mu}_{p}^{(d)}=\frac{\sigma_{\textrm{best-fit}}}{\sigma_{p,\, SM}^{(d)}}\,,
\]
where $d$ denotes the inclusive or exclusive final state searched for,
and $p=\textrm{inclusive, }$$GF$, $VBF$, $AP$ denotes the production
mode of the SM Higgs.

For some final states, as for instance $\gamma\gamma$, the experiments 
find excesses even compared to the expectations from a SM Higgs. These 
excesses are not statistically significant, but they nonetheless 
motivated further investigations of the 2011 and 2012 data in terms of 
various models of new physics. To increase their sensitivity to new 
physics both ATLAS and CMS performed a search for Higgs-like resonances 
in \emph{exclusive} final states.

To isolate exclusive final states the dedicated Higgs searches require 
tags of special kinds of events, such as the presence of two energetic 
forward jet with large rapidity separation and very little hadronic 
activity in the central part of the detector. This configuration of QCD 
radiation accompanying the Higgs is typical of the VBF production mode, and
therefore the ``dijet'' tag searches are essentially probes of the VBF 
production of the Higgs. Furthermore the experiments presented searches 
for the Higgs boson produced in association with a gauge boson, that are 
sensitive to different kinematical regions, but probe the same couplings 
as in VBF.

\bigskip

All the data presented by the experiments for Higgs searches can be 
translated into searches for the dilaton. In what follows we shall study 
how well the various dilaton scenarios fit the current data. We shall 
also try to see if in the current data one can spot the first hints of a 
preference for some of the scenarios of compositeness and EWSB described 
above. 

As reference values for the SM Higgs properties we take the central 
values reported in 
\cite{LHC-Higgs-Cross-Section-Working-Group:2011fj,CrossSectionsLHC,The-CDF-Collaboration:2012qy} 
and summarized in Tables \ref{tab:Higgs125} and \ref{tab:Higgs1265}. For the convenience of the 
reader and to complete the description of our input data, in Table 
\ref{tab:Signal-strength-best-fits} we collect the best-fit signal 
strength $\bar{\mu}_{p,d}\pm\delta\bar{\mu}_{p,d}$ used in our analysis.

\begin{table}
\begin{centering}
\begin{tabular}{|c|c|}
\hline 
\multicolumn{2}{|c|}{$BR(h\to XX)$}\tabularnewline
\hline 
\hline 
$bb$ & 0.58\tabularnewline
\hline 
$WW$ & 0.216\tabularnewline
\hline 
$gg$ & 0.085\tabularnewline
\hline 
$\tau\tau$ & 0.064\tabularnewline
\hline 
$c\bar{c}$ & 0.0291\tabularnewline
\hline 
$ZZ$ & 0.027\tabularnewline
\hline 
$\gamma\gamma$ & 0.00228\tabularnewline
\hline 
$Z\gamma$ & 0.00154\tabularnewline
\hline 
$\mu\bar{\mu}$ & 0.00022\tabularnewline
\hline 
\end{tabular}$\quad$%
\begin{tabular}{|c|c|c|c|}
\hline 
\multicolumn{4}{|c|}{$\sigma$ {[}pb{]}}\tabularnewline
\hline 
\hline
& LHC 8 TeV & LHC 7 TeV & TeVatron 1.96 TeV\tabularnewline
\hline 
GF & 19.5  & 15.3  & 0.949\tabularnewline
\hline 
VBF & 1.56 & 1.21 & 0.0653\tabularnewline
\hline 
WH &  0.697  & 0.573 & 0.129\tabularnewline
\hline 
ZH & 0.394  & 0.316 & 0.0785\tabularnewline 
\hline 
\end{tabular}
\par\end{centering}

\caption{\label{tab:Higgs125}Properties of the SM Higgs boson at 125 GeV from
\cite{LHC-Higgs-Cross-Section-Working-Group:2011fj,CrossSectionsLHC,The-CDF-Collaboration:2012qy}. }
\end{table}

\begin{table}
\begin{centering}
\begin{tabular}{|c|c|}
\hline 
\multicolumn{2}{|c|}{$BR(h\to XX)$}\tabularnewline
\hline 
\hline 
$bb$ & 0.553\tabularnewline
\hline 
$WW$ & 0.239\tabularnewline
\hline 
$gg$ & 0.0842\tabularnewline
\hline 
$\tau\tau$ & 0.0608\tabularnewline
\hline 
$c\bar{c}$ & 0.0279\tabularnewline
\hline 
$ZZ$ & 0.0302\tabularnewline
\hline 
$\gamma\gamma$ & 0.00228\tabularnewline
\hline 
$Z\gamma$ & 0.00163\tabularnewline
\hline 
$\mu\bar{\mu}$ & 0.000211\tabularnewline
\hline 
\end{tabular}$\quad$%
\begin{tabular}{|c|c|c|}
\hline 
\multicolumn{3}{|c|}{$\sigma$ {[}pb{]}}\tabularnewline
\hline 
\hline
& LHC 8 TeV & LHC 7 TeV \tabularnewline
\hline 
GF & 19.07  & 14.96   \tabularnewline
\hline 
VBF & 1.54 & 1.19 \tabularnewline
\hline 
WH &  0.669  & 0.550  \tabularnewline
\hline 
ZH & 0.379  & 0.304  \tabularnewline 
\hline 
\end{tabular}
\par\end{centering}

\caption{\label{tab:Higgs1265}Properties of the SM Higgs boson at 126.5 GeV from
\cite{LHC-Higgs-Cross-Section-Working-Group:2011fj,CrossSectionsLHC}. }
\end{table}

\bigskip

\begin{table}
\[
\begin{array}{cc}
 \text{ATLAS $\tau \tau $ 2011} & \text{0.2 $\pm $ 1.8} \\
 \text{ATLAS bb AP 2011} & \text{0.5 $\pm $ 2.} \\
 \text{ATLAS WW 2011+2012} & \text{1.4 $\pm $ 0.5} \\
 \text{ATLAS ZZ 2011+2012} & \text{1.3 $\pm $ 0.6} \\
 \text{ATLAS $\gamma \gamma $ 2011+2012} & \text{1.4 $\pm $ 0.5} \\
  \text{ATLAS $\gamma \gamma $ dijet 2011} & \text{2.7 $\pm $ 1.8} \\
  \text{ATLAS $\gamma \gamma $ dijet 2012} & \text{2.6 $\pm $ 1.7} \\
 \text{CMS WW VBF 2011+2012} & \text{0.2 $\pm $ 1.5} \\
 \text{CMS $\gamma \gamma $ Dijet Loose 2012} & \text{-0.6 $\pm $ 2.} \\
 \text{CMS $\gamma \gamma $ Dijet Tight 2012} & \text{1.3 $\pm $ 1.6} \\
 \text{CMS $\gamma \gamma $ Dijet 2011} & \text{4.2 $\pm $ 2.} \\
 \text{CMS WW AP 2011} & \text{-2.8 $\pm $ 3.} \\
 \text{CMS $\tau \tau $ 2011+2012} & \text{-0.2 $\pm $ 0.8} \\
 \text{CMS bb AP 2011+2012} & \text{0.5 $\pm $ 0.8} \\
 \text{CMS WW 2011+2012} & \text{0.6 $\pm $ 0.4} \\
 \text{CMS ZZ 2011+2012} & \text{0.8 $\pm $ 0.4} \\
 \text{CMS $\gamma \gamma $ 2011+2012} & \text{1.35 $\pm $ 0.44} \\
 \text{TeVatron $\gamma \gamma $} & \text{3.6 $\pm $ 2.8} \\
 \text{TeVatron WW} & \text{0.3 $\pm $ 1.1} \\
 \text{TeVatron bb AP} & \text{2. $\pm $ 0.7}
\end{array}
\]

\caption{\label{tab:Signal-strength-best-fits}Signal strength best-fits extracted
from \cite{ATLAS-CONF-2012-093,CMS-PAS-HIG-12-020,The-CDF-Collaboration:2012qy,CMS-PAS-HIG-12-015,CMS-PAS-HIG-12-008,ATLAS-Collaboration:2012fv,ATLAS-CONF-2012-091,ATLAS-CONF-2012-092,ATLAS-CONF-2012-098}.
 AP stands for associated production.
We  take the best-fit signal strength for $m_{h}=125$
GeV throughout but for the $\gamma\gamma$ searches of ATLAS where we take values at $m_{h}$=126.5 GeV.}
\end{table}

When not specified the searches have been interpreted as inclusive 
searches, dominated by the GF production mechanism, but with the VBF and 
AP subdominant contributions taken into account. For both ATLAS and CMS 
data the signal strengths for the inclusive $\gamma\gamma$ final state 
are computed combining in quadrature the relevant untagged di-photon 
categories. We checked that for a global combination of all the the 
$\gamma\gamma$ data the combination in quadrature reproduces the 
combination from the experiments within 10-20\%. The associated 
production searches are explicitly marked as AP and are considered as 
pure associated production. The VBF searches have different degree of 
contamination from the GF process with emissions of extra-jets. In our 
analysis the VBF searches are all in the $\gamma\gamma$ channel and we 
use the gluon fusion contamination reported in 
\cite{CMS-PAS-HIG-12-015,ATLAS-CONF-2012-091}. We discard the 
information about searches in exclusive channels where we were not able 
to reconstruct the gluon-fusion contamination from the the public data. 
This is the case of the ``CMS WW VBF 2011+2012'' search reported by CMS.

As a check of the reliability of our treatment of the experimental data 
we reproduced the analysis of CMS for a Beyond the Standard Model (BSM) 
scenario with modified Higgs couplings to fermions and vectors. We 
observe rather good agreement with the result shown in 
\cite{CMS-PAS-HIG-12-020} in the plane $\left( c_{V},\,c_{F}\right)$ 
(see Ref.~\cite{CMS-PAS-HIG-12-020} for a definition of these 
couplings).

\subsection{Results}

In this section we shall examine the 2011 and 2012 data from ATLAS, CMS 
and the TeVatron to assess whether the discovered boson at about 125 GeV is 
compatible with the properties of the general dilaton described above. 
In particular we shall determine what the current data tells us about 
the dynamics that underlies EWSB and the nature of the SM fermions in a 
dilaton scenario. We shall also discuss what observations can be done to 
cross-check the predictions of the different dilaton frameworks.

For each scenario we shall compute the minimum over the relevant parameters
of the $\chi^{2}$ between the observed data best-fit signal strengths
$\bar{\mu}_{p,d}\pm\delta\bar{\mu}_{p,d}$ of Table~\ref{tab:Signal-strength-best-fits}
and theory predictions $\mu_{p,d}$ for the the signal strengths 
\begin{equation}
\chi^{2}=\sum_{i=\{(p,d)\}}\frac{(\bar{\mu}_{i}-\mu_{i})^{2}}{\delta\bar{\mu}_{i}^{2}}\,,
\end{equation}
where $i$ runs over all the combinations of production modes $p$
and decays $d$ listed in Table \ref{tab:Signal-strength-best-fits}.

As a reference for the analysis of the dilaton scenarios we compute
the $\chi^{2}$ for the SM Higgs boson. We obtain $\chi^{2}=16.9$.
As we have a total of 19 measurements the SM seems in reasonable
agreement with the current data.

\subsection{Elementary fermions }

In what follows we shall perform a fit of the available data using the 
coupling structure of a scenario with the elementary top. While we shall 
consider a fairly large range of $\epsilon$, our primary focus will be 
on the region $0.35 \leq \epsilon \leq 0.55$. The upper limit on the 
preferred range originates from the need to suppress potentially 
dangerous effects in flavor physics with a scale not too far below 1000 
TeV, while the lower limit arises from the requirement that the CFT is a 
strongly coupled theory with significant anomalous dimensions, so that 
electroweak symmetry breaking is not overly fine tuned.

For the parameter $\xi$ we restrict the best-fit to lie in the range  
$0\leq\xi\leq2$. This range is suitable for a generic scenario that 
encompasses both $\xi$ values typical of a pNGB-like and 
technicolor-like mechanism for EWSB. The best fit parameters for this 
scenario are
 \begin{equation}
\xi=0.86,\,\epsilon=0.,\,\phi=1.8,\,\label{eq:EtopCFT}
\end{equation}
which yields $\chi^{2}=12.9$. Imposing $\epsilon\geq0.35$ we get $\chi^{2}=16.9$ for a best fit point $$\xi=0.68,\,\epsilon=0.35,\,\phi=2.2\,.$$ As we have 3 parameters for the couplings 
and a total of 19 measurements, the elementary top scenario seems in 
reasonable agreement with the current data for either choice of the minimal $\epsilon$.

Considering separately the measurements of ATLAS and CMS we obtain
a result very similar to the global one for CMS, while ATLAS tends to prefer larger $\xi$, due to the larger rates in $WW$ and $ZZ$ modes.

\begin{figure}
\begin{centering}
\includegraphics[width=0.55\linewidth]{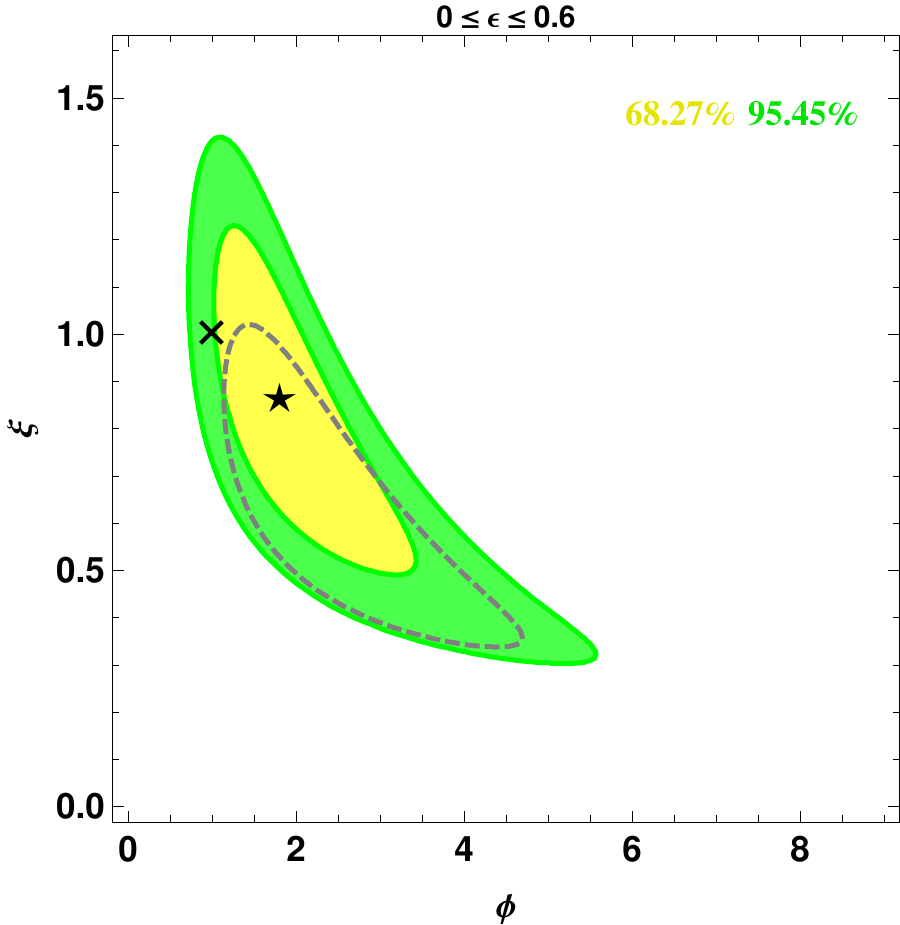}
\par\end{centering}

\caption{\label{fig:phi-xi-CFT-and-pNGB} The $1\sigma$ (yellow) and $2\sigma$(green)
CL regions of the plane $(\phi,\xi)$. At each point the $\chi^{2}$
has been minimized w.r.t. $\epsilon$ in the range $0\leq \epsilon \leq 0.6$. The black star corresponds to the
best-fit point and the black cross is the location of the SM-Higgs-like
dilaton. The dashed line corresponds to the $2\sigma$ contour applying the further constraint $\epsilon\geq0.35$}
\end{figure}

To visualize the global constraint on $\xi$ in Figure 
\ref{fig:phi-xi-CFT-and-pNGB} we show the the $1\sigma$ and $2\sigma$ CL 
regions of the $\chi^{2}$ minimized w.r.t $\epsilon$ in the plane 
$(\phi,\,\xi)\,$. The current data prefers rather large values of $\xi$, in 
the domain of technicolor-like models. However pNGB-like values of 
$\xi\sim0.3$ are still allowed within $2\sigma$.

\bigskip

We remark that in the minimization on $\epsilon$ performed in 
Figure~\ref{fig:phi-xi-CFT-and-pNGB} the value of $\epsilon$ tends to 
hit the lower boundary of the imposed range. As 
explained in the discussion that lead to 
Eq.~(\ref{eq:epsilonSensitive}), the value of $\epsilon\,$ is largely 
determined by the rates of the VBF and GF production modes. In the 
current data the most precise measurements are those in $\gamma\gamma$ 
and the GF-dominated channels do not exhibit a larger excess than VBF. 
As a consequence $\epsilon$ is driven to the minimal allowed value.

\begin{figure}
\begin{centering}
\includegraphics[width=0.55\linewidth]{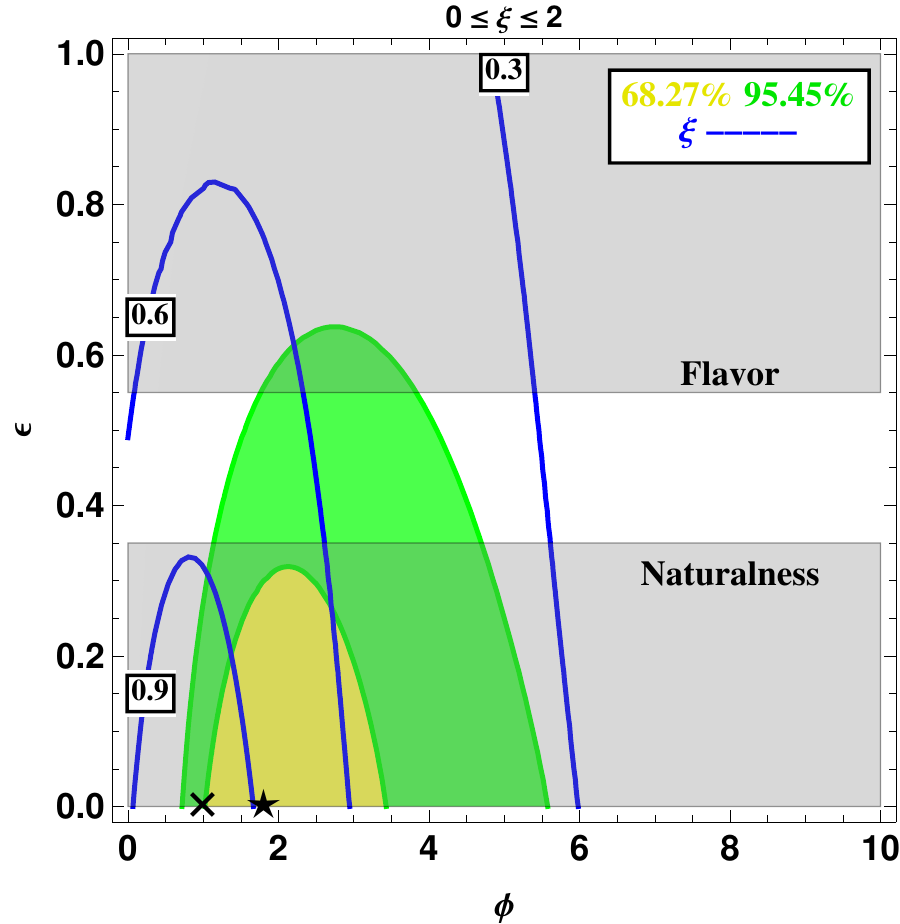}

\par\end{centering}

\caption{\label{fig:phi-epsilon-CFT}The $1\sigma$ (yellow) and $2\sigma$ (green) CL
regions of the $(\phi,\,\epsilon)$ plane. In each point the $\chi^{2}$
is minimized w.r.t $\xi$ in the interval $0\leq\xi\leq2$. The solid
blue lines are the isolines of the value of $\xi$ that minimizes
the $\chi^{2}$. The 
black star corresponds to the best-fit point and the 
 black cross is the location of the SM-Higgs-like dilaton. The upper gray-shaded region is disfavored by constraints from flavor physics. The lower gray-shaded region is disfavored by the bounds on the dimension of the operator $\mathcal{H}^{\dagger}\mathcal{H}$ and their implications for the solution of the hierarchy problem discussed in Section~\ref{fermioncouplings}.}

\end{figure}

The strength of this pull can be visualized in 
Figure~\ref{fig:phi-epsilon-CFT} where we show the $1\sigma$ and 
$2\sigma$ CL regions of the plane $(\phi,\,\epsilon)$. The figure shows 
that in the region corresponding to large $\xi$ points with big values 
of $\epsilon$, beyond $\epsilon=0.4$, are still allowed within the the 
$1\sigma$ CL region. In the pNGB-like region, that approximately lies 
close to and to the right of the isoline $\xi=0.3$, we remark that the 
current data tends to push $\epsilon$ towards zero, implying that this 
scenario is finetuned. It will be interesting to see how these 
constraints on $\epsilon$ will evolve when further data becomes 
available, especially on the ratio 
$\sigma_{VBF}^{(\gamma\gamma)}/\sigma_{GF}^{(\gamma\gamma)}$.

\subsection{Composite fermions}

As in the case of the elementary top scenario for the parameter $\xi$ we again consider  a generic range  $0\leq\xi\leq 2$. The best fit parameters for this scenario
are
\[
\xi=1.2,\,\psi=0.59,\,\phi=1.87\,,
\]
which gives $\chi^{2}=11.9\,$. The composite top scenario seems in a 
reasonable agreement with the current data. Just as in the case of 
elementary top we observe that ATLAS data prefers larger values of 
$\xi$.

We remark that the preferred values of $\psi$ are somewhat low. In 
particular at the best fit point the coupling of the gluons to the 
dilaton is suppressed with respect to the coupling of the Higgs in the 
SM. This is rather at odds with what we would expect from a CFT that 
contains colored states.

The tendency of the data to push $\psi$ toward low values can be
understood through Eq.~(\ref{eq:psiSensitive}). Analogously to what
happens for $\epsilon$ in the elementary top scenario, $\psi$ is
mostly determined by the ratio of VBF and GF rates and, due to the
precision of the data, the $\gamma\gamma$ final state has the stronger
pull. 

\begin{figure}
\begin{centering}
\includegraphics[width=0.55\linewidth]{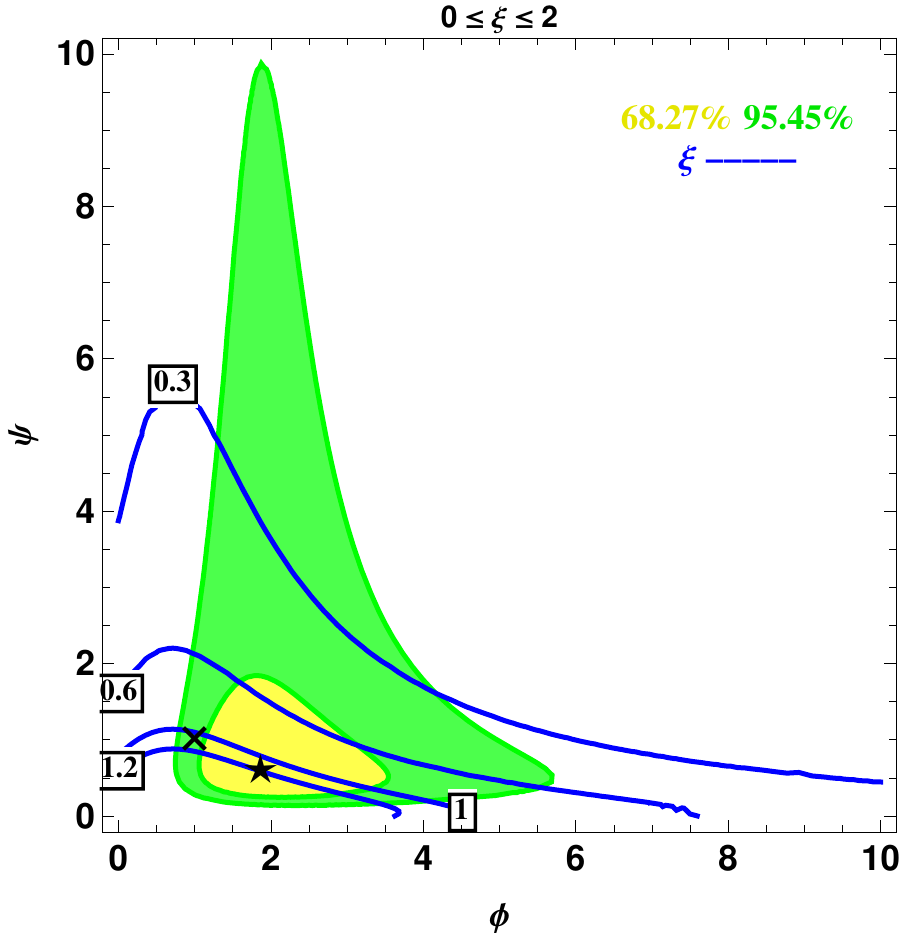}
\par\end{centering}

\caption{\label{fig:phi-psi-CFT-composite}The $1\sigma$  (yellow) and $2\sigma$ (green)
CL regions of the $(\phi,\,\psi)$ plane. In each point the $\chi^{2}$
is minimized w.r.t $\xi$ in the interval $0\leq\xi\leq2$. The solid
blue lines are the isolines of the value of $\xi$ that minimizes
the $\chi^{2}$. The black star corresponds to the best-fit point. The black cross is the location of
the SM-Higgs-like dilaton.}
\end{figure}

The strength of the pull on $\psi$ is displayed in 
Figure~\ref{fig:phi-psi-CFT-composite} where we show the $1\sigma$ and 
$2\sigma$ CL regions of the plane $(\phi,\,\psi)$ under the generic 
scenario for EWSB $0\leq\xi\leq2$. The figure shows that with small 
variations in $\xi$ it is possible to have $\psi\gtrsim1$ within the 
$2\sigma$ CL region indicated by the current data. We remark that the 
values of $\psi$ and $\phi$ in Eq.~(\ref{psi0phi0}) for the scenario of 
composite massless gauge boson lie outside of the $2\sigma$ CL region of 
the fit in our three parameter space. For completeness we report that 
the best fit for the scenario of composite massless gauge boson is 
$\xi=0.07$ which corresponds to $\chi^{2}=21.6$.
 
\begin{figure}
\begin{centering}
\includegraphics[width=0.55\linewidth]{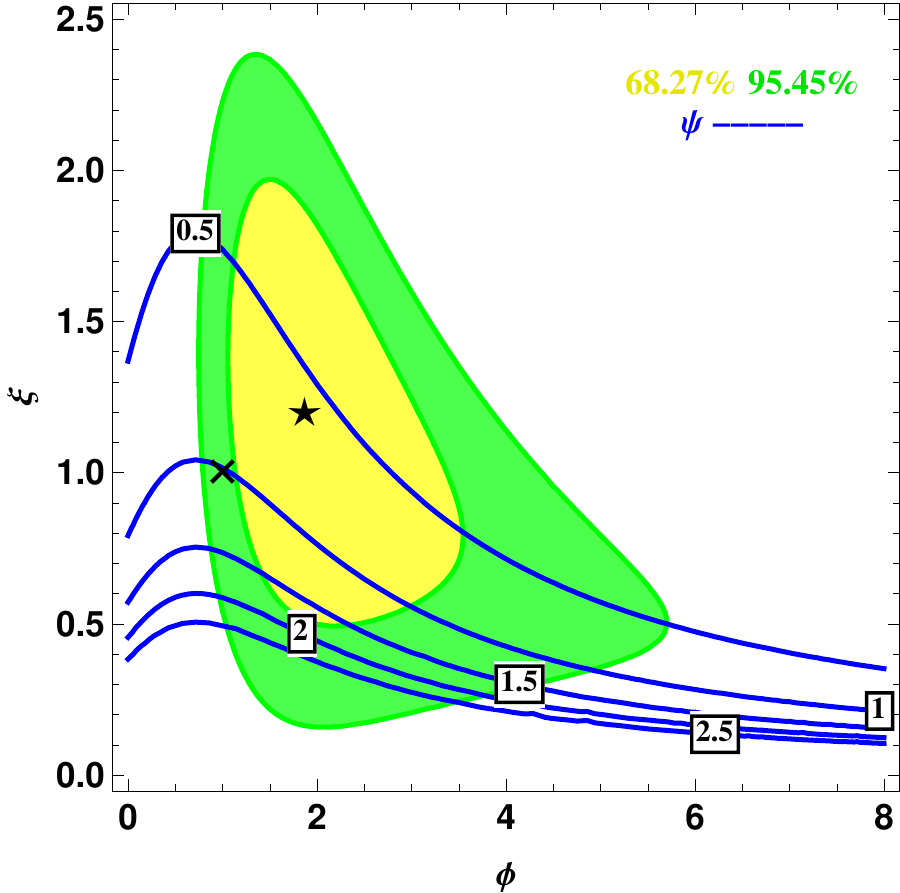}
\par\end{centering}

\caption{\label{fig:phi-xi-CFT-composite}The $1\sigma$ (yellow) and $2\sigma$(green)
CL regions of the $(\phi,\,\xi)$ plane. In each point the $\chi^{2}$
is minimized w.r.t $\psi$. The solid blue lines are the isolines of the
value of $\psi$ that minimizes the $\chi^{2}$. The black star corresponds
to the best-fit point. The black cross
is the location of the SM-Higgs-like dilaton.}
\end{figure}

Furthermore we show in Figure~\ref{fig:phi-xi-CFT-composite} the $1\sigma$ and $2\sigma$ CL regions in the $(\phi,\,\xi)$
plane together with the contours for the best-fit value of $\psi$. This figure makes explicit that reducing $\xi$ from its best-fit value it is possible to to accommodate points in the composite fermion parameter space with $\psi\gtrsim 1$ within the $2\sigma$ CL region.

\section{Conclusions}

In this paper we have studied in detail the possibility that the 125~GeV 
resonance observed at CERN is the dilaton associated with the 
spontaneous breaking of an approximate conformal symmetry. Our focus has 
been on theories where this conformal dynamics underlies the breaking of 
electroweak symmetry. The discussion has been mostly carried on in a 
language suitable to describe at the same time scenarios where there is 
a pNGB Higgs boson that is relatively heavy and therefore yet to be 
discovered, as well as scenarios where the breaking of the EW symmetry 
arises directly from strong dynamics as in technicolor models.

We have considered a rather wide spectrum of possibilities for the 
properties of the known SM states within the dilaton framework. In 
particular, we have allowed for both elementary and  
composite fermions. Furthermore we have considered both the case when
the SM gauge bosons are composites of the CFT and the case when they 
are elementary states.

We have discussed how the LHC Higgs data can be used to investigate the 
realization of all these scenarios for the dilaton. In particular we 
have introduced an effective description of the coupling structure that 
is suitable to take into account the leading order physics of the 
interactions of the dilaton with the SM states. The adopted 
parametrization naturally accommodates the 
effects of the breaking of conformal symmetry on the couplings of the 
dilaton. The summary of the coupling structure we have considered is 
reported in Table \ref{tab:couplings}. We stress that this 
parametrization exhibits wide coverage; for instance it captures dilatons 
from RS constructions with different embeddings of the fermions as well 
as dilatons for which there is no known corresponding RS realization.

Motivated by the impressive results of the experiments at the LHC, we 
have studied the specific measurements that are critical to uncovering 
the nature of the dilaton that could be responsible for the signal at 
125~GeV. For this it was essential to exploit the capabilities that the 
experiments have demonstrated in disentangling the several exclusive 
channels that contribute to the data about the 125~GeV resonance. We 
have shown that using this information it is also possible to 
address questions about the elementary or composite nature of the 
SM fermions and gauge bosons.

We have confronted the several possible incarnations of the dilaton 
against the set of available measurements in Higgs searches. We have 
identified the regions of the parameter space that are favored by the 
current measurements, and find that it is generically preferred to have 
a dilaton whose couplings resemble rather closely those of SM Higgs 
boson. Although we find that the coupling structure of the dilaton does 
in principle allow better agreement with the central values of the 
current measurements than the SM Higgs, the uncertainties are at the 
moment sufficiently large that from a global $\chi^{2}$ analysis a 
statistically significant preference does not emerge. Since further data 
from the 2012 run of the LHC may begin to highlight significant 
deviations from the predictions of a SM Higgs, we believe that it is 
worth considering the dilaton, in its full generality, as a potential 
explanation of the new physics at 125~GeV.

\section*{Acknowledgements}

We are glad to thank Christophe Grojean and David Poland for discussions 
and comments on our manuscript. RF also thanks Roberto Contino and 
Alessandro Strumia for important discussions. This work is supported by 
the NSF under grant PHY-0968854. The work of RF is supported by the NSF 
Grants PHY-0910467 and by the Maryland Center for Fundamental Physics. 
RF thanks CERN TH division for hospitality and support while this work 
was completed. RF acknowledges the hospitality of the Aspen Center for 
Physics, which is supported by the NSF Grant No. PHY-1066293.

\bigskip

{\bf Note added:} While we were completing this paper, we were informed of the upcoming work~\cite{Bellazzini:2012fr} which overlaps with some of the ideas presented here.

\clearpage

\providecommand{\href}[2]{#2}\begingroup\raggedright\endgroup


\begin{thebibliography}{10}

\bibitem{The-CMS-Collaboration:2012ys}
{The CMS Collaboration}, ``{Observation of a new boson at a mass of 125 GeV
  with the CMS experiment at the LHC},'' {\em ArXiv e-prints} (July, 2012) ,
  \href{http://arxiv.org/abs/1207.7235}{{\ttfamily arXiv:1207.7235 [hep-ex]}}.
  \url{https://twiki.cern.ch/twiki/bin/view/CMSPublic/Hig12028TWiki}.

\bibitem{The-ATLAS-Collaboration:2012rt}
{The ATLAS Collaboration}, ``{Observation of a new particle in the search for
  the Standard Model Higgs boson with the ATLAS detector at the LHC},'' {\em
  ArXiv e-prints} (July, 2012) ,
  \href{http://arxiv.org/abs/1207.7214}{{\ttfamily arXiv:1207.7214 [hep-ex]}}.
  \url{https://atlas.web.cern.ch/Atlas/GROUPS/PHYSICS/PAPERS/HIGG-2012-27/}.

\bibitem{Carmi:2012yq}
D.~{Carmi}, A.~{Falkowski}, E.~{Kuflik}, and T.~{Volansky}, ``{Interpreting LHC
  Higgs Results from Natural New Physics Perspective},'' {\em ArXiv e-prints}
  (Feb., 2012) , \href{http://arxiv.org/abs/1202.3144}{{\ttfamily
  arXiv:1202.3144 [hep-ph]}}.

\bibitem{Azatov:2012rt}
A.~{Azatov}, R.~{Contino}, and J.~{Galloway}, ``{Model-Independent Bounds on a
  Light Higgs},'' {\em ArXiv e-prints} (Feb., 2012) ,
  \href{http://arxiv.org/abs/1202.3415}{{\ttfamily arXiv:1202.3415 [hep-ph]}}.

\bibitem{Espinosa:2012vn}
J.~R. {Espinosa}, C.~{Grojean}, M.~{Muhlleitner}, and M.~{Trott},
  ``{Fingerprinting Higgs Suspects at the LHC},'' {\em ArXiv e-prints} (Feb.,
  2012) , \href{http://arxiv.org/abs/1202.3697}{{\ttfamily arXiv:1202.3697
  [hep-ph]}}.

\bibitem{Giardino:2012lr}
P.~P. {Giardino}, K.~{Kannike}, M.~{Raidal}, and A.~{Strumia},
  ``{Reconstructing Higgs boson properties from the LHC and Tevatron data},''
  {\em ArXiv e-prints} (Mar., 2012) ,
  \href{http://arxiv.org/abs/1203.4254}{{\ttfamily arXiv:1203.4254 [hep-ph]}}.

\bibitem{Ellis:2012dq}
J.~{Ellis} and T.~{You}, ``{Global Analysis of Experimental Constraints on a
  Possible Higgs-Like Particle with Mass \~{} 125 GeV},'' {\em ArXiv e-prints}
  (Apr., 2012) , \href{http://arxiv.org/abs/1204.0464}{{\ttfamily
  arXiv:1204.0464 [hep-ph]}}.

\bibitem{Klute:2012yq}
M.~{Klute}, R.~{Lafaye}, T.~{Plehn}, M.~{Rauch}, and D.~{Zerwas}, ``{Measuring
  Higgs Couplings from LHC Data},'' {\em ArXiv e-prints} (May, 2012) ,
  \href{http://arxiv.org/abs/1205.2699}{{\ttfamily arXiv:1205.2699 [hep-ph]}}.

\bibitem{Corbett:2012rt}
T.~{Corbett}, O.~J.~P. {Eboli}, J.~{Gonzalez-Fraile}, and M.~C.
  {Gonzalez-Garcia}, ``{Constraining anomalous Higgs interactions},'' {\em
  ArXiv e-prints} (July, 2012) ,
  \href{http://arxiv.org/abs/1207.1344}{{\ttfamily arXiv:1207.1344 [hep-ph]}}.

\bibitem{Giardino:2012qy}
P.~P. {Giardino}, K.~{Kannike}, M.~{Raidal}, and A.~{Strumia}, ``{Is the
  resonance at 125 GeV the Higgs boson?},'' {\em ArXiv e-prints} (July, 2012) ,
  \href{http://arxiv.org/abs/1207.1347}{{\ttfamily arXiv:1207.1347 [hep-ph]}}.

\bibitem{Carmi:2012vn}
D.~{Carmi}, A.~{Falkowski}, E.~{Kuflik}, T.~{Volansky}, and J.~{Zupan},
  ``{Higgs After the Discovery: A Status Report},'' {\em ArXiv e-prints} (July,
  2012) , \href{http://arxiv.org/abs/1207.1718}{{\ttfamily arXiv:1207.1718
  [hep-ph]}}.

\bibitem{Espinosa:2012yq}
J.~R. {Espinosa}, C.~{Grojean}, M.~{Muhlleitner}, and M.~{Trott}, ``{First
  Glimpses at Higgs' face},'' {\em ArXiv e-prints} (July, 2012) ,
  \href{http://arxiv.org/abs/1207.1717}{{\ttfamily arXiv:1207.1717 [hep-ph]}}.

\bibitem{Ellis:2012fj}
J.~{Ellis} and T.~{You}, ``{Global Analysis of the Higgs Candidate with Mass
  \~{} 125 GeV},'' {\em ArXiv e-prints} (July, 2012) ,
  \href{http://arxiv.org/abs/1207.1693}{{\ttfamily arXiv:1207.1693 [hep-ph]}}.

\bibitem{Low:2012ys}
I.~{Low}, J.~{Lykken}, and G.~{Shaughnessy}, ``{Have We Observed the Higgs
  (Imposter)?},'' {\em ArXiv e-prints} (July, 2012) ,
  \href{http://arxiv.org/abs/1207.1093}{{\ttfamily arXiv:1207.1093 [hep-ph]}}.

\bibitem{Montull:2012kx}
M.~{Montull} and F.~{Riva}, ``{Higgs discovery: the beginning or the end of
  natural EWSB?},'' {\em ArXiv e-prints} (July, 2012) ,
  \href{http://arxiv.org/abs/1207.1716}{{\ttfamily arXiv:1207.1716 [hep-ph]}}.

\bibitem{Buckley:2012fk}
M.~R. {Buckley} and D.~{Hooper}, ``{Are There Hints of Light Stops in Recent
  Higgs Search Results?},'' {\em ArXiv e-prints} (July, 2012) ,
  \href{http://arxiv.org/abs/1207.1445}{{\ttfamily arXiv:1207.1445 [hep-ph]}}.

\bibitem{Gunion:2012uq}
J.~F. {Gunion}, Y.~{Jiang}, and S.~{Kraml}, ``{Could two NMSSM Higgs bosons be
  present near 125 GeV?},'' {\em ArXiv e-prints} (July, 2012) ,
  \href{http://arxiv.org/abs/1207.1545}{{\ttfamily arXiv:1207.1545 [hep-ph]}}.

\bibitem{Gabrielli:2012rt}
E.~{Gabrielli}, K.~{Kannike}, B.~{Mele}, A.~{Racioppi}, and M.~{Raidal},
  ``{Fermiophobic Higgs boson and supersymmetry},'' {\em ArXiv e-prints} (Mar.,
  2012) , \href{http://arxiv.org/abs/1204.0080}{{\ttfamily arXiv:1204.0080
  [hep-ph]}}.

\bibitem{Gabrielli:2012vn}
E.~{Gabrielli}, B.~{Mele}, and M.~{Raidal}, ``{Has a fermiophobic Higgs boson
  been detected at the LHC?},'' {\em ArXiv e-prints} (Feb., 2012) ,
  \href{http://arxiv.org/abs/1202.1796}{{\ttfamily arXiv:1202.1796 [hep-ph]}}.

\bibitem{Espinosa:2012qy}
J.~R. {Espinosa}, M.~{Muhlleitner}, C.~{Grojean}, and M.~{Trott}, ``{Probing
  for Invisible Higgs Decays with Global Fits},'' {\em ArXiv e-prints} (May,
  2012) , \href{http://arxiv.org/abs/1205.6790}{{\ttfamily arXiv:1205.6790
  [hep-ph]}}.

\bibitem{Dreiner:2012lr}
H.~K. {Dreiner}, J.~S. {Kim}, and O.~{Lebedev}, ``{First LHC Constraints on
  Neutralinos},'' {\em ArXiv e-prints} (June, 2012) ,
  \href{http://arxiv.org/abs/1206.3096}{{\ttfamily arXiv:1206.3096 [hep-ph]}}.

\bibitem{Azatov:2012ys}
A.~{Azatov}, R.~{Contino}, D.~{Del Re}, J.~{Galloway}, M.~{Grassi}, and
  S.~{Rahatlou}, ``{Determining Higgs couplings with a model-independent
  analysis of $h \to \gamma \gamma$},'' {\em ArXiv e-prints} (Apr., 2012) ,
  \href{http://arxiv.org/abs/1204.4817}{{\ttfamily arXiv:1204.4817 [hep-ph]}}.

\bibitem{Kraml:2012fr}
{Kraml, S. et al.}, ``{Searches for new physics: Les Houches recommendations
  for the presentation of LHC results},''
  \href{http://dx.doi.org/10.1140/epjc/s10052-012-1976-3}{{\em European
  Physical Journal C} {\bfseries 72} (Apr., 2012) 1976},
  \href{http://arxiv.org/abs/1203.2489}{{\ttfamily arXiv:1203.2489 [hep-ph]}}.

\bibitem{Susskind:1978ms}
L.~Susskind, ``{Dynamics of Spontaneous Symmetry Breaking in the Weinberg-Salam
  Theory},''
\href{http://dx.doi.org/10.1103/PhysRevD.20.2619}{{\em Phys.Rev.} {\bfseries
  D20} (1979) 2619--2625}.

\bibitem{Weinberg:1975gm}
S.~Weinberg, ``{Implications of Dynamical Symmetry Breaking},''
\href{http://dx.doi.org/10.1103/PhysRevD.13.974}{{\em Phys.Rev.} {\bfseries
  D13} (1976) 974--996}.

\bibitem{Hill:2002ap}
C.~T. Hill and E.~H. Simmons, ``{Strong dynamics and electroweak symmetry
  breaking},'' \href{http://dx.doi.org/10.1016/S0370-1573(03)00140-6}{{\em
  Phys.Rept.} {\bfseries 381} (2003) 235--402},
\href{http://arxiv.org/abs/hep-ph/0203079}{{\ttfamily arXiv:hep-ph/0203079
  [hep-ph]}}.

\bibitem{Georgi:1974yw}
H.~Georgi and A.~Pais, ``{Calculability and Naturalness in Gauge Theories},''
\href{http://dx.doi.org/10.1103/PhysRevD.10.539}{{\em Phys.Rev.} {\bfseries
  D10} (1974) 539}.

\bibitem{Georgi:1975tz}
H.~Georgi and A.~Pais, ``{Vacuum Symmetry and the PseudoGoldstone
  Phenomenon},''
\href{http://dx.doi.org/10.1103/PhysRevD.12.508}{{\em Phys.Rev.} {\bfseries
  D12} (1975) 508}.

\bibitem{Kaplan:1983fs}
D.~B. Kaplan and H.~Georgi, ``{SU(2) x U(1) Breaking by Vacuum Misalignment},''
\href{http://dx.doi.org/10.1016/0370-2693(84)91177-8}{{\em Phys.Lett.}
  {\bfseries B136} (1984) 183}.

\bibitem{Kaplan:1983sm}
D.~B. Kaplan, H.~Georgi, and S.~Dimopoulos, ``{Composite Higgs Scalars},''
\href{http://dx.doi.org/10.1016/0370-2693(84)91178-X}{{\em Phys.Lett.}
  {\bfseries B136} (1984) 187}.

\bibitem{Georgi:1984af}
H.~Georgi and D.~B. Kaplan, ``{Composite Higgs and Custodial SU(2)},''
\href{http://dx.doi.org/10.1016/0370-2693(84)90341-1}{{\em Phys.Lett.}
  {\bfseries B145} (1984) 216}.

\bibitem{Luty:2006sp}
M.~A. {Luty} and T.~{Okui}, ``{Conformal technicolor},''
  \href{http://dx.doi.org/10.1088/1126-6708/2006/09/070}{{\em Journal of High
  Energy Physics} {\bfseries 9} (Sept., 2006) 70},
  \href{http://arxiv.org/abs/arXiv:hep-ph/0409274}{{\ttfamily
  arXiv:hep-ph/0409274}}.

\bibitem{Holdom:1984sk}
B.~Holdom, ``{Techniodor},''
\href{http://dx.doi.org/10.1016/0370-2693(85)91015-9}{{\em Phys.Lett.}
  {\bfseries B150} (1985) 301}.

\bibitem{Appelquist:1986an}
T.~W. Appelquist, D.~Karabali, and L.~Wijewardhana, ``{Chiral Hierarchies and
  the Flavor Changing Neutral Current Problem in Technicolor},''
\href{http://dx.doi.org/10.1103/PhysRevLett.57.957}{{\em Phys.Rev.Lett.}
  {\bfseries 57} (1986) 957}.

\bibitem{Yamawaki:1985zg}
K.~Yamawaki, M.~Bando, and K.-i. Matumoto, ``{Scale Invariant Technicolor Model
  and a Technidilaton},''
\href{http://dx.doi.org/10.1103/PhysRevLett.56.1335}{{\em Phys.Rev.Lett.}
  {\bfseries 56} (1986) 1335}.

\bibitem{Appelquist:1986tr}
T.~Appelquist and L.~Wijewardhana, ``{Chiral Hierarchies and Chiral
  Perturbations in Technicolor},''
\href{http://dx.doi.org/10.1103/PhysRevD.35.774}{{\em Phys.Rev.} {\bfseries
  D35} (1987) 774}.

\bibitem{Appelquist:1987fc}
T.~Appelquist and L.~Wijewardhana, ``{Chiral Hierarchies from Slowly Running
  Couplings in Technicolor Theories},''
\href{http://dx.doi.org/10.1103/PhysRevD.36.568}{{\em Phys.Rev.} {\bfseries
  D36} (1987) 568}.

\bibitem{Salam:1970qk}
A.~Salam and J.~Strathdee, ``{Nonlinear realizations. 2. Conformal symmetry},''
\href{http://dx.doi.org/10.1103/PhysRev.184.1760}{{\em Phys.Rev.} {\bfseries
  184} (1969) 1760--1768}.

\bibitem{Isham:1971dv}
C.~Isham, A.~Salam, and J.~Strathdee, ``{Nonlinear realizations of space-time
  symmetries. Scalar and tensor gravity},''
\href{http://dx.doi.org/10.1016/0003-4916(71)90269-7}{{\em Annals Phys.}
  {\bfseries 62} (1971) 98--119}.

\bibitem{Zuminolectures}
{B. Zumino}, ``{Lectures on Elementary Particles and Quantum Field Theory,
  edited by S.~Deser {\it et. al.}, MIT Press (1970)},'' {\em {1970 Brandeis
  Summer School}} .

\bibitem{Ellis:1970yd}
J.~R. Ellis, ``{Aspects of conformal symmetry and chirality},''
\href{http://dx.doi.org/10.1016/0550-3213(70)90422-0}{{\em Nucl.Phys.}
  {\bfseries B22} (1970) 478--492}.

\bibitem{Rattazzi:2000hs}
R.~{Rattazzi} and A.~{Zaffaroni}, ``{Comments on the Holographic Picture of the
  Randall-Sundrum Model},''
  \href{http://dx.doi.org/10.1088/1126-6708/2001/04/021}{{\em Journal of High
  Energy Physics} {\bfseries 4} (Apr., 2001) 21},
  \href{http://arxiv.org/abs/arXiv:hep-th/0012248}{{\ttfamily
  arXiv:hep-th/0012248}}.

\bibitem{Goldberger:2007zk}
{Goldberger, W.D. and Grinstein, B. and Skiba, W.}, ``{Distinguishing the Higgs
  Boson from the Dilaton at the Large Hadron Collider},''
  \href{http://dx.doi.org/10.1103/PhysRevLett.100.111802}{{\em Physical Review
  Letters} {\bfseries 100} no.~11, (Mar., 2008) 111802},
  \href{http://arxiv.org/abs/0708.1463}{{\ttfamily arXiv:0708.1463 [hep-ph]}}.

\bibitem{Vecchi:2010gj}
L.~Vecchi, ``{Phenomenology of a light scalar: the dilaton},''
  \href{http://dx.doi.org/10.1103/PhysRevD.82.076009}{{\em Phys.Rev.}
  {\bfseries D82} (2010) 076009},
\href{http://arxiv.org/abs/1002.1721}{{\ttfamily arXiv:1002.1721 [hep-ph]}}.

\bibitem{Chacko:2012sy}
{Z.~Chacko  and R.~K.~Mishra}, ``{Effective theory of a light dilaton},'' 
\href{http://arxiv.org/abs/1209.3022}{{\ttfamily arXiv:1209.3022 [hep-ph]}}.

\bibitem{Fan:2008jk}
J.~Fan, W.~D. Goldberger, A.~Ross, and W.~Skiba, ``{Standard Model couplings
  and collider signatures of a light scalar},''
  \href{http://dx.doi.org/10.1103/PhysRevD.79.035017}{{\em Phys.Rev.}
  {\bfseries D79} (2009) 035017},
\href{http://arxiv.org/abs/0803.2040}{{\ttfamily arXiv:0803.2040 [hep-ph]}}.

\bibitem{Coleppa:2012wd}
B.~{Coleppa}, T.~{Gr{\'e}goire}, and H.~E. {Logan}, ``{Dilaton constraints and
  LHC prospects},'' \href{http://dx.doi.org/10.1103/PhysRevD.85.055001}{{\em
  Phys. Rev. D} {\bfseries 85} no.~5, (Mar., 2012) 055001},
  \href{http://arxiv.org/abs/1111.3276}{{\ttfamily arXiv:1111.3276 [hep-ph]}}.

\bibitem{Campbell:2012lr}
B.~A. {Campbell}, J.~{Ellis}, and K.~A. {Olive}, ``{Phenomenology and cosmology
  of an electroweak pseudo-dilaton and electroweak baryons},''
  \href{http://dx.doi.org/10.1007/JHEP03(2012)026}{{\em Journal of High Energy
  Physics} {\bfseries 3} (Mar., 2012) 26},
  \href{http://arxiv.org/abs/1111.4495}{{\ttfamily arXiv:1111.4495 [hep-ph]}}.

\bibitem{de-Sandes:2012gf}
H.~{de Sandes} and R.~{Rosenfeld}, ``{Radion-Higgs mixing effects on bounds
  from LHC Higgs boson searches},''
  \href{http://dx.doi.org/10.1103/PhysRevD.85.053003}{{\em Phys. Rev. D}
  {\bfseries 85} no.~5, (Mar., 2012) 053003},
  \href{http://arxiv.org/abs/1111.2006}{{\ttfamily arXiv:1111.2006 [hep-ph]}}.

\bibitem{Barger:2012qf}
V.~{Barger}, M.~{Ishida}, and W.-Y. {Keung}, ``{Differentiating the Higgs Boson
  from the Dilaton and Radion at Hadron Colliders},''
  \href{http://dx.doi.org/10.1103/PhysRevLett.108.101802}{{\em Physical Review
  Letters} {\bfseries 108} no.~10, (Mar., 2012) 101802},
  \href{http://arxiv.org/abs/1111.4473}{{\ttfamily arXiv:1111.4473 [hep-ph]}}.

\bibitem{Barger:2012ve}
V.~{Barger}, M.~{Ishida}, and W.-Y. {Keung}, ``{Dilaton at the LHC},''
  \href{http://dx.doi.org/10.1103/PhysRevD.85.015024}{{\em Phys. Rev. D}
  {\bfseries 85} no.~1, (Jan., 2012) 015024},
  \href{http://arxiv.org/abs/1111.2580}{{\ttfamily arXiv:1111.2580 [hep-ph]}}.

\bibitem{Elander:2012fv}
D.~{Elander} and M.~{Piai}, ``{The decay constant of the holographic
  techni-dilaton and the 125 GeV boson},'' {\em ArXiv e-prints} (Aug., 2012) ,
  \href{http://arxiv.org/abs/1208.0546}{{\ttfamily arXiv:1208.0546 [hep-ph]}}.

\bibitem{Matsuzaki:2012rp}
S.~{Matsuzaki} and K.~{Yamawaki}, ``{Is 125 GeV techni-dilaton found at
  LHC?},'' {\em ArXiv e-prints} (July, 2012) ,
  \href{http://arxiv.org/abs/1207.5911}{{\ttfamily arXiv:1207.5911 [hep-ph]}}.

\bibitem{Matsuzaki:2012fr}
S.~{Matsuzaki} and K.~{Yamawaki}, ``{Discovering 125 GeV techni-dilaton at
  LHC},'' {\em ArXiv e-prints} (June, 2012) ,
  \href{http://arxiv.org/abs/1206.6703}{{\ttfamily arXiv:1206.6703 [hep-ph]}}.

\bibitem{Matsuzaki:2012pd}
S.~{Matsuzaki} and K.~{Yamawaki}, ``{Techni-dilaton at 125 GeV},'' {\em ArXiv
  e-prints} (Jan., 2012) , \href{http://arxiv.org/abs/1201.4722}{{\ttfamily
  arXiv:1201.4722 [hep-ph]}}.
  
\bibitem{Matsuzaki:2012pi}
S.~{Matsuzaki} and K.~{Yamawaki}, ``{Holographic techni-dilaton at 125 GeV},''
  {\em ArXiv e-prints} (Sept., 2012) ,
  \href{http://arxiv.org/abs/1209.2017}{{\ttfamily arXiv:1209.2017 [hep-ph]}}.

\bibitem{Grzadkowski:2012bh}
B.~{Grzadkowski}, J.~F. {Gunion}, and M.~{Toharia}, ``{Higgs-radion
  interpretation of the LHC data?},''
  \href{http://dx.doi.org/10.1016/j.physletb.2012.04.037}{{\em Physics Letters
  B} {\bfseries 712} (May, 2012) 70--80},
  \href{http://arxiv.org/abs/1202.5017}{{\ttfamily arXiv:1202.5017 [hep-ph]}}.

\bibitem{Cheung:2012lq}
K.~{Cheung} and T.-C. {Yuan}, ``{Could the Excess Seen at 124-126 GeV Be due to
  the Randall-Sundrum Radion?},''
  \href{http://dx.doi.org/10.1103/PhysRevLett.108.141602}{{\em Physical Review
  Letters} {\bfseries 108} no.~14, (Apr., 2012) 141602},
  \href{http://arxiv.org/abs/1112.4146}{{\ttfamily arXiv:1112.4146 [hep-ph]}}.

\bibitem{Maldacena:1997re}
J.~{Maldacena}, ``{The Large-N Limit of Superconformal Field Theories and
  Supergravity},'' \href{http://dx.doi.org/10.1023/A:1026654312961}{{\em
  International Journal of Theoretical Physics} {\bfseries 38} (1999)
  1113--1133}, \href{http://arxiv.org/abs/arXiv:hep-th/9711200}{{\ttfamily
  arXiv:hep-th/9711200}}.

\bibitem{Witten:1998qj}
E.~{Witten}, ``{Anti De Sitter Space And Holography},'' {\em ArXiv High Energy
  Physics - Theory e-prints} (Feb., 1998) ,
  \href{http://arxiv.org/abs/arXiv:hep-th/9802150}{{\ttfamily
  arXiv:hep-th/9802150}}.

\bibitem{Verlinde:1999fy}
H.~{Verlinde}, ``{Holography and compactification},''
  \href{http://dx.doi.org/10.1016/S0550-3213(00)00224-8}{{\em Nuclear Physics
  B} {\bfseries 580} (July, 2000) 264--274},
  \href{http://arxiv.org/abs/arXiv:hep-th/9906182}{{\ttfamily
  arXiv:hep-th/9906182}}.

\bibitem{Verlinde:1999xm}
E.~{Verlinde} and H.~{Verlinde}, ``{RG-flow, gravity and the cosmological
  constant},'' \href{http://dx.doi.org/10.1088/1126-6708/2000/05/034}{{\em
  Journal of High Energy Physics} {\bfseries 5} (May, 2000) 34},
  \href{http://arxiv.org/abs/arXiv:hep-th/9912018}{{\ttfamily
  arXiv:hep-th/9912018}}.

\bibitem{Randall:1999ee}
L.~{Randall} and R.~{Sundrum}, ``{Large Mass Hierarchy from a Small Extra
  Dimension},'' \href{http://dx.doi.org/10.1103/PhysRevLett.83.3370}{{\em
  Physical Review Letters} {\bfseries 83} (Oct., 1999) 3370--3373},
  \href{http://arxiv.org/abs/arXiv:hep-ph/9905221}{{\ttfamily
  arXiv:hep-ph/9905221}}.

\bibitem{Csaki:2003zu}
C.~{Cs{\'a}ki}, C.~{Grojean}, L.~{Pilo}, and J.~{Terning}, ``{Towards a
  Realistic Model of Higgsless Electroweak Symmetry Breaking},''
  \href{http://dx.doi.org/10.1103/PhysRevLett.92.101802}{{\em Physical Review
  Letters} {\bfseries 92} no.~10, (Mar., 2004) 101802},
  \href{http://arxiv.org/abs/arXiv:hep-ph/0308038}{{\ttfamily
  arXiv:hep-ph/0308038}}.

\bibitem{Contino:2003ve}
R.~{Contino}, Y.~{Nomura}, and A.~{Pomarol}, ``{Higgs as a holographic
  pseudo-Goldstone boson},''
  \href{http://dx.doi.org/10.1016/j.nuclphysb.2003.08.027}{{\em Nuclear Physics
  B} {\bfseries 671} (Nov., 2003) 148--174},
  \href{http://arxiv.org/abs/arXiv:hep-ph/0306259}{{\ttfamily
  arXiv:hep-ph/0306259}}.

\bibitem{Agashe:2004rs}
K.~{Agashe}, R.~{Contino}, and A.~{Pomarol}, ``{The minimal composite Higgs
  model},'' \href{http://dx.doi.org/10.1016/j.nuclphysb.2005.04.035}{{\em
  Nuclear Physics B} {\bfseries 719} (July, 2005) 165--187},
  \href{http://arxiv.org/abs/arXiv:hep-ph/0412089}{{\ttfamily
  arXiv:hep-ph/0412089}}.

\bibitem{Goldberger:1999uk}
W.~D. Goldberger and M.~B. Wise, ``{Modulus stabilization with bulk fields},''
  \href{http://dx.doi.org/10.1103/PhysRevLett.83.4922}{{\em Phys.Rev.Lett.}
  {\bfseries 83} (1999) 4922--4925},
\href{http://arxiv.org/abs/hep-ph/9907447}{{\ttfamily arXiv:hep-ph/9907447
  [hep-ph]}}.

\bibitem{Csaki:2000lr}
C.~{Csaki}, M.~{Graesser}, L.~{Randall}, and J.~{Terning}, ``{Cosmology of
  brane models with radion stabilization},''
  \href{http://dx.doi.org/10.1103/PhysRevD.62.045015}{{\em Phys. Rev. D}
  {\bfseries 62} no.~4, (Aug., 2000) 045015},
  \href{http://arxiv.org/abs/arXiv:hep-ph/9911406}{{\ttfamily
  arXiv:hep-ph/9911406}}.

\bibitem{Goldberger:1999un}
W.~D. Goldberger and M.~B. Wise, ``{Phenomenology of a stabilized modulus},''
  \href{http://dx.doi.org/10.1016/S0370-2693(00)00099-X}{{\em Phys.Lett.}
  {\bfseries B475} (2000) 275--279},
\href{http://arxiv.org/abs/hep-ph/9911457}{{\ttfamily arXiv:hep-ph/9911457
  [hep-ph]}}.

\bibitem{Giudice:2000av}
G.~F. {Giudice}, R.~{Rattazzi}, and J.~D. {Wells}, ``{Graviscalars from
  higher-dimensional metrics and curvature-Higgs mixing},''
  \href{http://dx.doi.org/10.1016/S0550-3213(00)00686-6}{{\em Nuclear Physics
  B} {\bfseries 595} (Feb., 2001) 250--276},
  \href{http://arxiv.org/abs/arXiv:hep-ph/0002178}{{\ttfamily
  arXiv:hep-ph/0002178}}.

\bibitem{Csaki:2000zn}
C.~Csaki, M.~L. Graesser, and G.~D. Kribs, ``{Radion dynamics and electroweak
  physics},'' \href{http://dx.doi.org/10.1103/PhysRevD.63.065002}{{\em
  Phys.Rev.} {\bfseries D63} (2001) 065002},
\href{http://arxiv.org/abs/hep-th/0008151}{{\ttfamily arXiv:hep-th/0008151
  [hep-th]}}.

\bibitem{Rizzo:2002pq}
T.~G. Rizzo, ``{Radion couplings to bulk fields in the Randall-Sundrum
  model},'' {\em JHEP} {\bfseries 0206} (2002) 056,
\href{http://arxiv.org/abs/hep-ph/0205242}{{\ttfamily arXiv:hep-ph/0205242
  [hep-ph]}}.

\bibitem{Csaki:2007ns}
C.~Csaki, J.~Hubisz, and S.~J. Lee, ``{Radion phenomenology in realistic warped
  space models},'' \href{http://dx.doi.org/10.1103/PhysRevD.76.125015}{{\em
  Phys.Rev.} {\bfseries D76} (2007) 125015},
\href{http://arxiv.org/abs/0705.3844}{{\ttfamily arXiv:0705.3844 [hep-ph]}}.

\bibitem{Isham:1970gz}
C.~Isham, A.~Salam, and J.~Strathdee, ``{Spontaneous breakdown of conformal
  symmetry},''
\href{http://dx.doi.org/10.1016/0370-2693(70)90177-2}{{\em Phys.Lett.}
  {\bfseries B31} (1970) 300--302}.

\bibitem{Ellis:1971sa}
J.~R. Ellis, ``{Phenomenological actions for spontaneously-broken conformal
  symmetry},''
\href{http://dx.doi.org/10.1016/0550-3213(71)90193-3}{{\em Nucl.Phys.}
  {\bfseries B26} (1971) 536--546}.

\bibitem{RattazziPlanck2010}
R.~Rattazzi, ``The naturally light dilaton,'' {\em Talk at Planck 2010} .

\bibitem{Appelquist:2010gy}
T.~Appelquist and Y.~Bai, ``{A Light Dilaton in Walking Gauge Theories},''
  \href{http://dx.doi.org/10.1103/PhysRevD.82.071701}{{\em Phys.Rev.}
  {\bfseries D82} (2010) 071701},
\href{http://arxiv.org/abs/1006.4375}{{\ttfamily arXiv:1006.4375 [hep-ph]}}.

\bibitem{Tang:2012yq}
Y.~{Tang}, ``{Implications of LHC Searches for Massive Graviton},'' {\em ArXiv
  e-prints} (June, 2012) , \href{http://arxiv.org/abs/1206.6949}{{\ttfamily
  arXiv:1206.6949 [hep-ph]}}.

\bibitem{Tang:2012ly}
Y.~Tang, ``Comment on "could the excess seen at 124-126 gev be due to the
  randall-sundrum radion?", {\em ArXiv
  e-prints} (April, 2012) ,
  \href{http://arxiv.org/abs/1204.6145v1}{{\ttfamily arXiv:1204.6145 [hep-ph]}}.


\bibitem{Rattazzi:2008pe}
R.~{Rattazzi}, V.~S. {Rychkov}, E.~{Tonni}, and A.~{Vichi}, ``{Bounding scalar
  operator dimensions in 4D CFT},''
  \href{http://dx.doi.org/10.1088/1126-6708/2008/12/031}{{\em Journal of High
  Energy Physics} {\bfseries 12} (Dec., 2008) 31},
  \href{http://arxiv.org/abs/0807.0004}{{\ttfamily arXiv:0807.0004 [hep-th]}}.

\bibitem{Rychkov:2009ij}
V.~S. Rychkov and A.~Vichi, ``{Universal Constraints on Conformal Operator
  Dimensions},'' \href{http://dx.doi.org/10.1103/PhysRevD.80.045006}{{\em
  Phys.Rev.} {\bfseries D80} (2009) 045006},
\href{http://arxiv.org/abs/0905.2211}{{\ttfamily arXiv:0905.2211 [hep-th]}}.

\bibitem{Rattazzi:2010yc}
R.~{Rattazzi}, S.~{Rychkov}, and A.~{Vichi}, ``{Bounds in 4D conformal field
  theories with global symmetry},''
  \href{http://dx.doi.org/10.1088/1751-8113/44/3/035402}{{\em Journal of
  Physics A Mathematical General} {\bfseries 44} no.~3, (Jan., 2011) 035402},
  \href{http://arxiv.org/abs/1009.5985}{{\ttfamily arXiv:1009.5985 [hep-th]}}.

\bibitem{Vichi:2011ux} 
A.~{Vichi}, ``{Improved bounds for CFT's with global symmetries},''
  \href{http://dx.doi.org/10.1007/JHEP01(2012)162}{{\em Journal of High Energy
  Physics} {\bfseries 1} (Jan., 2012) 162},
  \href{http://arxiv.org/abs/1106.4037}{{\ttfamily arXiv:1106.4037 [hep-th]}}.

 
\bibitem{Poland:2011ey}
D.~Poland, D.~Simmons-Duffin, and A.~Vichi, ``{Carving Out the Space of 4D
  CFTs},'' \href{http://dx.doi.org/10.1007/JHEP05(2012)110}{{\em JHEP}
  {\bfseries 1205} (2012) 110},
\href{http://arxiv.org/abs/1109.5176}{{\ttfamily arXiv:1109.5176 [hep-th]}}.
 

\bibitem{Poland:2010wg}
D.~{Poland} and D.~{Simmons-Duffin}, ``{Bounds on 4D conformal and
  superconformal field theories},''
  \href{http://dx.doi.org/10.1007/JHEP05(2011)017}{{\em Journal of High Energy
  Physics} {\bfseries 5} (May, 2011) 17},
  \href{http://arxiv.org/abs/1009.2087}{{\ttfamily arXiv:1009.2087 [hep-th]}}.

\bibitem{Kaplan:1991dc}
D.~B. Kaplan, ``{Flavor at SSC energies: A New mechanism for dynamically
  generated fermion masses},''
\href{http://dx.doi.org/10.1016/S0550-3213(05)80021-5}{{\em Nucl.Phys.}
  {\bfseries B365} (1991) 259--278}.

\bibitem{Contino:2006nn}
R.~Contino, T.~Kramer, M.~Son, and R.~Sundrum, ``{Warped/composite
  phenomenology simplified},''
  \href{http://dx.doi.org/10.1088/1126-6708/2007/05/074}{{\em JHEP} {\bfseries
  0705} (2007) 074},
\href{http://arxiv.org/abs/hep-ph/0612180}{{\ttfamily arXiv:hep-ph/0612180
  [hep-ph]}}.

\bibitem{ATLAS-CONF-2012-093}
{ATLAS Collaboration}, ``{Observation of an Excess of Events in the Search for
  the Standard Model Higgs boson with the ATLAS detector at the LHC},'' Tech.
  Rep. ATLAS-CONF-2012-093, CERN, Geneva, Jul, 2012.
\newblock
  \url{https://atlas.web.cern.ch/Atlas/GROUPS/PHYSICS/CONFNOTES/ATLAS-CONF-201%
2-093/}.

\bibitem{CMS-PAS-HIG-12-020}
{CMS Collaboration}, ``{Observation of a new boson with a mass near 125 GeV},''
  {\em CMS-PAS-HIG-12-020} (2012) .
  \url{https://twiki.cern.ch/twiki/bin/view/CMSPublic/Hig12020TWiki}.

\bibitem{LHC-Higgs-Cross-Section-Working-Group:2011fj}
{LHC Higgs Cross Section Working Group}, ``{Handbook of LHC Higgs Cross
  Sections: 1. Inclusive Observables},'' {\em eprint arXiv:1101.0593} (Jan.,
  2011) , \href{http://arxiv.org/abs/1101.0593}{{\ttfamily arXiv:1101.0593
  [hep-ph]}}.

\bibitem{CrossSectionsLHC}
``{LHC Higgs Cross Section Working Group}.''
\newblock \url{https://twiki.cern.ch/twiki/bin/view/LHCPhysics/CrossSections}.

\bibitem{The-CDF-Collaboration:2012qy}
{The CDF Collaboration}, {the D0 Collaboration}, t.~{Tevatron New Physics}, and
  {Higgs Working Group}, ``{Updated Combination of CDF and D0 Searches for
  Standard Model Higgs Boson Production with up to 10.0 fb-1 of Data},'' {\em
  ArXiv e-prints} (July, 2012) ,
  \href{http://arxiv.org/abs/1207.0449}{{\ttfamily arXiv:1207.0449 [hep-ex]}}.

\bibitem{CMS-PAS-HIG-12-015}
{CMS Collaboration}, ``{Evidence for a new state decaying into two photons in
  the search for the standard model Higgs boson in pp collisions},'' {\em
  CMS-PAS-HIG-12-015} (2012) .
  \url{https://twiki.cern.ch/twiki/bin/view/CMSPublic/Hig12015TWiki}.

\bibitem{CMS-PAS-HIG-12-008}
{CMS Collaboration}, ``{Combination of SM, SM4, FP Higgs boson searches},''
  {\em CMS-PAS-HIG-12-008} (2012) .
  \url{https://twiki.cern.ch/twiki/bin/view/CMSPublic/Hig12008TWiki}.

\bibitem{ATLAS-Collaboration:2012fv}
{ATLAS Collaboration}, ``{Combined search for the Standard Model Higgs boson in
  pp collisions at $\sqrt{s}$ = 7 TeV with the ATLAS detector},'' {\em ArXiv
  e-prints} (July, 2012) , \href{http://arxiv.org/abs/1207.0319}{{\ttfamily
  arXiv:1207.0319 [hep-ex]}}.

\bibitem{ATLAS-CONF-2012-091}
``{Observation of an excess of events in the search for the Standard Model
  Higgs boson in the gamma-gamma channel with the ATLAS detector},'' Tech. Rep.
  ATLAS-CONF-2012-091, CERN, Geneva, Jul, 2012.
\newblock
  \url{https://atlas.web.cern.ch/Atlas/GROUPS/PHYSICS/CONFNOTES/ATLAS-CONF-201%
2-091/}.

\bibitem{ATLAS-CONF-2012-092}
{ATLAS Collaboration}, ``{Observation of an excess of events in the search for
  the Standard Model Higgs boson in the $H\rightarrow ZZ^{(*)}\rightarrow
  4\ell$ channel with the ATLAS detector},'' Tech. Rep. ATLAS-CONF-2012-092,
  CERN, Geneva, Jul, 2012.
\newblock
  \url{https://atlas.web.cern.ch/Atlas/GROUPS/PHYSICS/CONFNOTES/ATLAS-CONF-201%
2-092/}.

\bibitem{ATLAS-CONF-2012-098}
{ATLAS Collaboration}, ``{Observation of an Excess of Events in the Search for
  the Standard Model Higgs Boson in the $H\to WW^{(\ast)}\to \ell\nu\ell\nu$
  Channel with the ATLAS Detector},'' Tech. Rep. ATLAS-CONF-2012-098, CERN,
  Geneva, Jul, 2012.
\newblock
  \url{https://atlas.web.cern.ch/Atlas/GROUPS/PHYSICS/CONFNOTES/ATLAS-CONF-201%
2-098/}.

\bibitem{Bellazzini:2012fr}
B.~{Bellazzini}, C.~{Cs{\'a}ki}, J.~{Hubisz}, J.~{Serra}, and J.~{Terning},
  ``{A Higgslike Dilaton},'' {\em ArXiv e-prints} (Sept., 2012) ,
  \href{http://arxiv.org/abs/1209.3299}{{\ttfamily arXiv:1209.3299 [hep-ph]}}.



\end{thebibliography}
\end{document}